\documentclass[aps,superscriptaddress,preprint]{revtex4-1}

\usepackage{lineno}

\usepackage{graphicx}
\usepackage{amsmath,amssymb}
\usepackage{color}
\usepackage{dcolumn}
\usepackage{bm}
\usepackage{float}
\usepackage{threeparttable}

\begin{document}

\title{Scale dependence of energy transfer in turbulent plasma}

\author{Yan Yang}
\affiliation{Southern University of Science and Technology, Shenzhen, Guangdong 518055, China}
\affiliation{University of Science and Technology of China, Hefei, Anhui 230026, China}
\author{Minping Wan}
\email{wanmp@sustc.edu.cn}
\affiliation{Southern University of Science and Technology, Shenzhen, Guangdong 518055, China}
\author{William H. Matthaeus}
\affiliation{University of Delaware, Newark, DE 19716, USA}
\author{Luca Sorriso-Valvo}
\affiliation{Nanotec/CNR, Sede di Cosenza, Ponte P. Bucci, Cubo 31C, 87036 Rende, Italy}
\author{Tulasi N. Parashar}
\affiliation{University of Delaware, Newark, DE 19716, USA}
\author{Quanming Lu}
\affiliation{University of Science and Technology of China, Hefei, Anhui 230026, China}
\author{Yipeng Shi}
\affiliation{Peking University, Beijing 100871, China}
\author{Shiyi Chen}
\affiliation{Southern University of Science and Technology, Shenzhen, Guangdong 518055, China}

\begin{abstract}
In the context of space and astrophysical plasma turbulence and particle heating,
several vocabularies emerge for estimating turbulent energy dissipation rate,
including Kolmogorov-Yaglom third-order law and, in its various forms,
$\boldsymbol{j}\cdot\boldsymbol{E}$ (work done by the electromagnetic field on particles),
and $-\left( \boldsymbol{P} \cdot \nabla \right) \cdot \boldsymbol{u}$ (pressure-strain interaction),
to name a couple.
It is now understood that these energy transfer channels,
to some extent, are correlated with coherent structures.
In particular, we find that different energy dissipation proxies,
although not point-wise correlated,
are concentrated in proximity to each other,
for which they decorrelate in a few $d_i$(s).
However, the energy dissipation proxies dominate at different scales.
For example, there is an inertial range over which the third-order law is meaningful.
Contributions from scale bands stemming from scale-dependent spatial filtering show that,
the energy exchange through $\boldsymbol{j}\cdot\boldsymbol{E}$ mainly results from large scales,
while the energy conversion from fluid flow to internal through
$-\left( \boldsymbol{P} \cdot \nabla \right) \cdot \boldsymbol{u}$
dominates at small scales.
\end{abstract}

\maketitle



\section{Introduction}

Energy dissipation mechanism for weakly collisional or collisionless plasma
is of principal importance for addressing long-standing puzzles like
the acceleration of energetic particles, and
related questions that arise in space and astrophysical applications such as
the solar wind.
Depending on the plasma conditions,
different dissipative mechanisms like wave-particle interactions
{\citep{Markovskii06, Hollweg86, Hollweg02, Gary03, Gary08,Howes2008kinetic,He2015evidence,He2015proton}} and
heating by coherent structures and reconnections
{\citep{Dmitruk04, Retino07, Sundkvist07, Parashar11, TenBarge13, Perri12,He2018plasma}},
might be dominant at kinetic scales.
This question is inherently related to an open debate as to
whether the fluctuations in the solar wind are interacting waves, nonlinearly evolving turbulence,
or their mutual competition, each having its adherents
\citep{Bale2005Measurement,GoldreichSridhar95,Narita2011Dispersion,Sahraoui09,Sahraoui2010Three,Saito2008whistler,
Smith2012Observational,Schekochihin2008gyrokinetic,Howes2008kinetic,Howes2011gyrokinetic}.
Here we focus on well-defined quantitative parameters that describe collisionless dissipation
without being sidetracked by these controversies.

While at small scales kinetic processes must be considered,
MHD model remains a credible approximation for a kinetic plasma at scales
large enough to be well separated from kinetic effects.
Therefore instead of studying specific mechanisms at kinetic scales,
one can invoke the classical turbulence theory at MHD scales.
A standard turbulence scenario inherited from hydrodynamics displays
an energy cascade process over the MHD inertial range.
In MHD, the energy cascade within the inertial range satisfies
the Politano-Pouquet law \citep{Politano98a}, that describes
the scaling law of the mixed third-order moment of Elsasser
fields increments.
Under suitable assumptions (i.e., isotropy, homogeneity, time stationarity, and incompressibility),
the third order law follows a linear scaling relation with
scale separation and is proportional to mean energy dissipation rate.
The Politano-Pouquet law has been examined in the solar wind
\citep{Valvo07,MacBride08,Marino2008Heating,Stawarz09,Coburn15,Bandyopadhyay2018solar}
and in numerical simulations
\citep{Valvo02,Sorriso2018local}.
A number of studies have also taken into account corrections from
anisotropy \citep{Osman11,Stawarz2011third,Wan2009third,Wan2010third,Podesta2008laws},
compressibility \citep{Hadid2017energy,Kritsuk09,Carbone09,Forman10,Valvo10,Marino11,Galtier13,Banerjee16b},
solar wind shear \citep{Wan2009third,Wan2010third}
and expansion \citep{Gogoberidze2013yaglom,Hellinger2013proton}
and Hall effect \citep{Galtier2008karman,Andres2018exact,Hellinger2018karman}, to name a few.
More recently, \citet{Bandyopadhyay2018incompressive}
found that the standard Kolmogorov cascade may be operative at a diminished
intensity even in the kinetic scales, as an ingredient of a complex cascade accommodating kinetic effects.
We avoid all such complications here in an effort to elucidate
possible correlations amongst several basic estimations of energy dissipation
rate (energy dissipation proxies).

``Dissipation'' in this paper simply refers to the increase in internal energy of distribution functions.
This increase in internal energy is eventually ``thermalized'' by infrequent collisions but
this irreversibility is not our focus here.
The work done by the electromagnetic field on particles, $\boldsymbol{j}\cdot \boldsymbol{E}$
\citep{Wan12,Wan15,Osman2015multi,Fu2017intermittent,Chasapis2018situ,Howes2018spatially,Ergun2018magnetic,Yao2017direct},
although not strictly irreversible dissipation,
is the necessary energy supply from electromagnetic fields that
ultimately goes into the internal energy reservoir.
It is therefore customary to deem $\boldsymbol{j}\cdot \boldsymbol{E}$ as a dissipation proxy,
which is highly localized in association with current sheets \citep{Wan12,Wan15,Osman2015multi,Chasapis2018situ}.
A basic question, however, is what fraction of the electromagnetic energy released
through $\boldsymbol{j}\cdot \boldsymbol{E}$ ends up as ion and electron random motion as opposed to
fluid flow. To this end, \citet{YangEA-PRE-17,YangEA-PoP-17} proposed recently
that the interaction between pressure tensor and strain tensor,
$-\left( \boldsymbol{P} \cdot \nabla \right) \cdot \boldsymbol{u}$,
is responsible for the generation of plasma internal energy.
This idea can be traced back to early works by \citet{Braginskii65},
which is generalized into the kinetic realm of collisionless systems
by Del Sarto et al. \citep{DelSarto16,DelSarto2017shear} as well.
More recently, there is growing evidence that elucidates the role of pressure-strain interaction
using numerical simulations \citep{Sitnov2018kinetic} and observations \citep{Chasapis2018energy}.

The work done so far has not specifically emphasized
the associations and differences that exist among these dissipation proxies,
yet simulations and observations indicate that each of them plays
an important role in the heating process.
Here we seek to
describe their correlations. To take it further,
they might differ from each other in many ways as well.
We adopt a narrow tack here,
inquiring at what scales the different proxies dominate,
thus providing more detail concerning energy transfer
from macroscopic fluid scales to kinetic scales.

\section{Simulation Details}

Here we employ a fully kinetic simulation by P3D \citep{Zeiler02}
in 2.5D geometry (three components of dependent
field vectors and a two-dimensional spatial grid).
Number density is normalized to a reference
number density $n_r$ (=1 in this simulation),
mass to proton mass $m_i$ (=1 in this simulation),
charge to proton charge $q_i$, and
magnetic field to a reference $B_r$ (=1 in this run).
Length is normalized to the ion inertial length $d_i$,
time to the ion cyclotron time $\Omega_i^{-1}$,
velocity to the reference Alfv{\'e}n speed
$v_{Ar}=B_r/\left(4\pi m_i n_r\right)^{1/2}$, and
temperature to $T_r=m_i v_{Ar}^2$.
The simulation was performed in a periodic domain, whose size
is $L=149.5648 d_i$, with $4096^2$ grid points and
$3200$ particles of each species
per cell ($\sim 107\times 10^9$ total particles).
The ion to electron mass ratio is $m_i/m_e = 25$,
and the speed of light in the simulation is $c=15 v_{Ar}$.
Although small mass ratio
and low speed of light might
introduce some unrealistic effects,
they are necessary compromises to
attain large simulated system size $L/d_i$, and reasonably
large particle number per cell
that also run for long dynamical times.
The run is
a decaying initial value problem,
starting with uniform density ($n_0=1.0$)
and temperature of ions and electrons ($T_0=0.3$).
The uniform magnetic field is $B_0 = 1.0$ directed
out of the plane.
This simulation is a part of a set of simulations to study kinetic
plasma turbulence as a function of plasma $\beta$ \citep{ParasharApJL18}.

\begin{figure}[!htpb]
\centering
\includegraphics[width=0.6\textwidth]{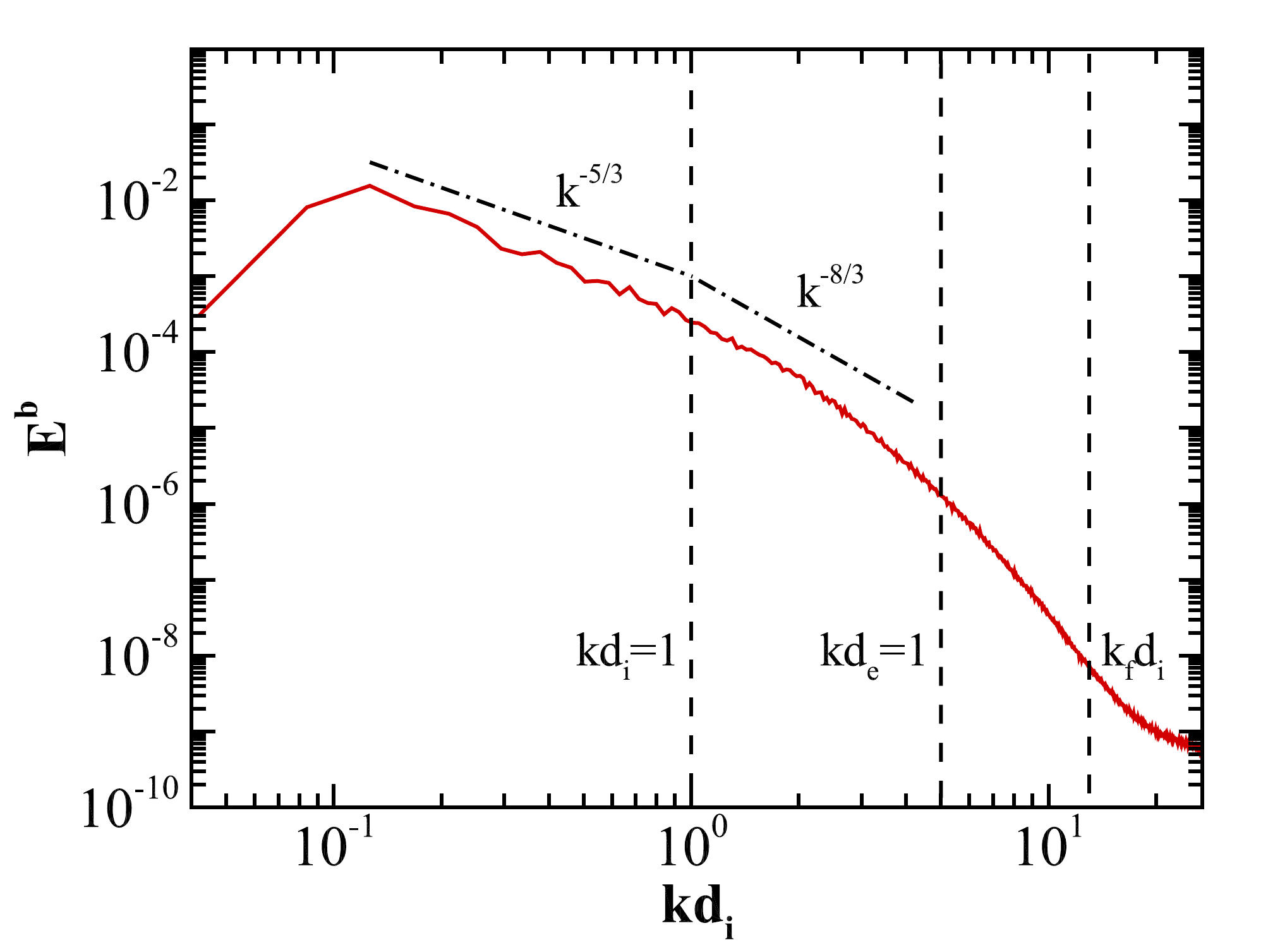}
\caption{Omnidirectional energy spectrum of magnetic
fluctuations (prior to filtering).
Power laws are shown for reference.
Vertical lines are drawn at wavenumbers corresponding to ion inertial scale,
to electron inertial scale, and at the filtering scale,
as explained in the text.}
\label{Fig.Eb-spectra}
\end{figure}

We analyze statistics using a snapshot near the time
of maximum root mean square (r.m.s.) electric current density
(i.e., $t\Omega_{i}=99.0$).
Fig.~\ref{Fig.Eb-spectra} shows the omnidirectional energy spectrum of magnetic fluctuations,
where $k^{-5/3}$ and $k^{-8/3}$ power laws are shown for reference in
the range $k<1/d_i$ and $1/d_i < k < 1/d_e$, respectively.
Prior to statistical analyses,
we remove noise inherent in the particle-in-cell plasma algorithm through
low-pass Fourier filtering of the fields
at $k_f d_i\sim 13$.

\section{Third-order Moment Estimate}

\begin{figure}[!htpb]
\centering
\includegraphics[width=0.6\textwidth]{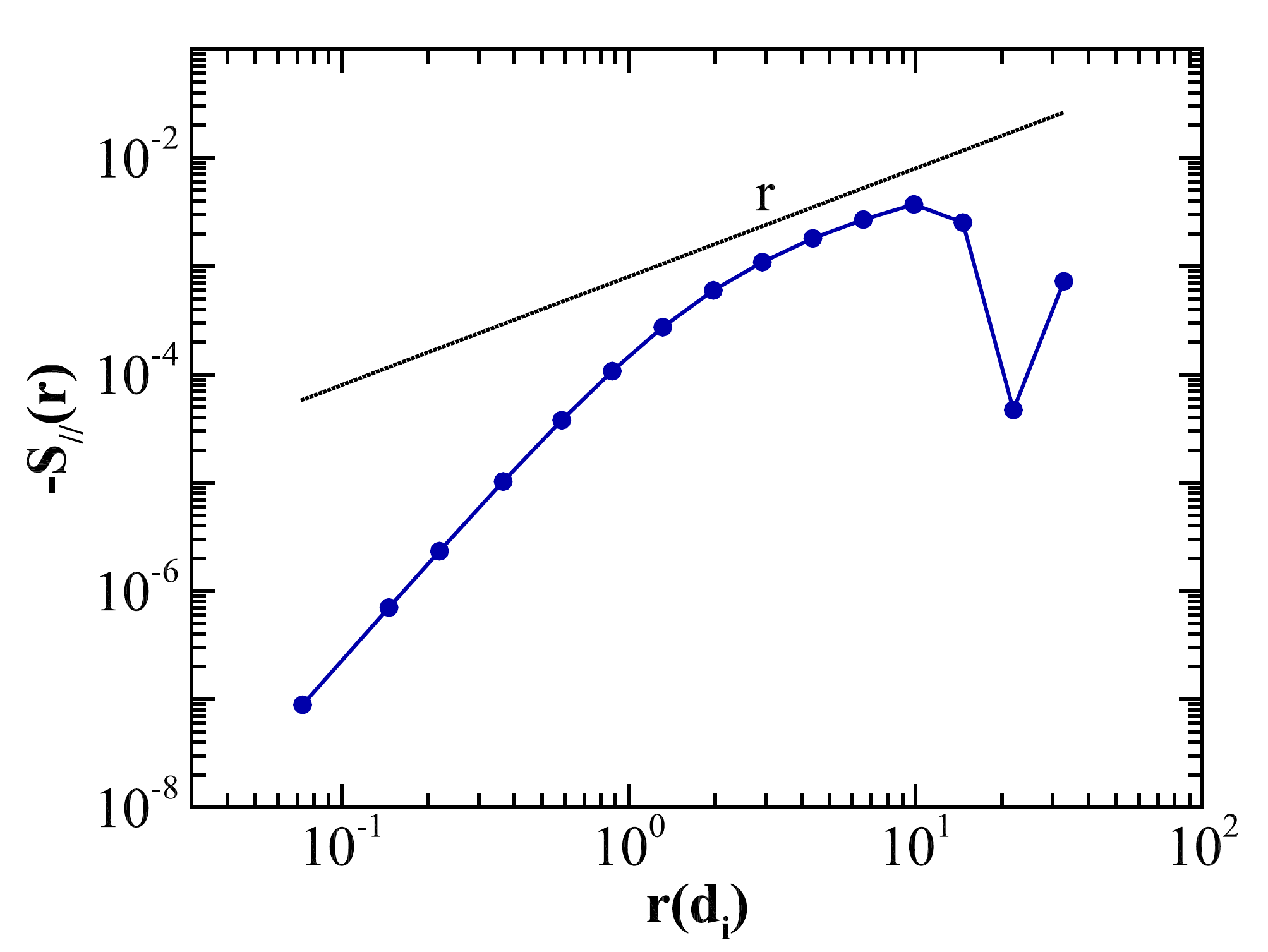}
\caption{Mixed third-order moment of Elsasser fields increments,
$S_{\parallel}=\left(S^{+}_{\parallel}+S^{-}_{\parallel}\right)/2$, as a function of separation length $r$.
A linear fit is also indicated.
The mean energy transfer rate evaluated within $\left[2d_i, 10d_i\right]$ range is
$\epsilon=1.87\pm 0.20 \ (\times 10^{-4}\ {v_{Ar}^3 d_i^{-1}})$.}
\label{Fig.Yaglom-scaling}
\end{figure}

In studying energy dissipation, the most satisfactory way to proceed in MHD would be
to directly study the viscous and resistive dissipation functions.
However, in default of an explicit expression for dissipation in collisionless plasma,
it is promising to appeal to a third-order law such as \citep{Politano98a},
\begin{equation}
S^{\pm}_{\parallel}(r) = \langle\delta z^{\mp}_{\parallel} \delta z^{\pm}_{i} \delta z^{\pm}_{i}\rangle = -{\frac{4}{d}} \epsilon^{\pm}{r}
\end{equation}
where $\boldsymbol{z}^{\pm}=\boldsymbol{u} \pm \boldsymbol{b}/\sqrt{4\pi \rho}$,
$\delta\boldsymbol{z}^{\pm}=\boldsymbol{z}^{\pm}\left(\boldsymbol{x}+\boldsymbol{r}\right)-\boldsymbol{z}^{\pm}\left(\boldsymbol{x}\right)$,
$\delta z^{\pm}_{\parallel}=\delta\boldsymbol{z}^{\pm} \cdot \boldsymbol{r}/{r}$,
$\epsilon^{\pm}$ is the mean energy transfer rate,
and $d$ is the spatial dimension ($d=2$ in our case).
Upon defining $S_{\parallel}=\left(S^{+}_{\parallel}+S^{-}_{\parallel}\right)/2$,
we arrive at an expression that
is proportional to the total energy (kinetic and magnetic energies) transfer
rate $\epsilon=\left(\epsilon^{+} + \epsilon^{-}\right)/2$ and the separation $r$.
We investigate the validity of aforementioned third-order law in Fig.~\ref{Fig.Yaglom-scaling},
which exhibits nearly a decade of range ($\sim \left[2d_i, 10d_i\right]$)
over which a linear variation with separation fits well,
thus finding an approximately constant energy transfer rate across this inertial range of scales.
The mean energy transfer rate evaluated within $\left[2d_i, 10d_i\right]$ range is
$\epsilon=1.87\pm 0.20 \ (\times 10^{-4}\ {v_{Ar}^3 d_i^{-1}})$.
However, we should also keep in mind that
the system does not necessarily realize an inertial range
that terminates exactly at the ion scale,
although the case at hand shows a break of the linear law in the vicinity of the ion scale.
If proceeding to smaller scales, e.g., sub-ion scales,
one would prefer a dynamic law that accommodates Hall effects \citep{Galtier2008karman,Andres2018exact,Hellinger2018karman,Bandyopadhyay2018incompressive},
a procedure that we will defer to a future study.
Here we also emphasize that a compressible channel for transfer also exists \citep{YangEA-POF-17},
but compressional effects are expected to be weak for the present quasi-incompressible turbulence
simulation (as established by the density fluctuation
$\delta \rho'_\alpha/\langle\rho_\alpha\rangle=
\sqrt{\langle(\rho_\alpha-\langle\rho_\alpha\rangle)^2\rangle}/\langle\rho_\alpha\rangle \sim 0.07$.)

The third-order law itself is averaged over space, but is well resolved in scale.
It is of interest to unravel the third order expression, examining the contributions
to the final result from each point in space.
Accordingly, following \citet{Sorriso2018statistical,Sorriso2018local},
we denote
by $\epsilon_r^{\pm}$ the ``local" pseudo-energy transfer rate at the scale $r$,
\begin{equation}
\epsilon_r^{\pm} = - {\frac{d}{4}}{\frac{\delta z^{\mp}_{\parallel} \delta z^{\pm}_{i} \delta z^{\pm}_{i}}{r}},
\end{equation}
so that the ``local" total (kinetic and magnetic) energy transfer
rate (LET) is computed as
\begin{equation}
\epsilon_r={\frac{\epsilon_r^+ + \epsilon_r^-}{2}}.
\end{equation}
The new measure, LET, is a scalar field spatially localized.
However, the third-order law is only valid in a statistical (or ensemble average) sense.
The LET defined in this way might be reminiscent of, but not equivalent to,
the scale-to-scale energy transfer.
In particular, caution is required in claiming that
the LET, whose sign can be positive or negative, reveals exactly
how much energy is transferred locally towards smaller or larger scales
in each point of the domain.
With these caveats in mind, we still believe that the LET
can help to identify patches with enhanced energy transfer,
as suggested in \citet{Marsch1997intermittency,Sorriso2018statistical}.

\begin{figure}[!htpb]
\centering
\includegraphics[width=1.0\textwidth]{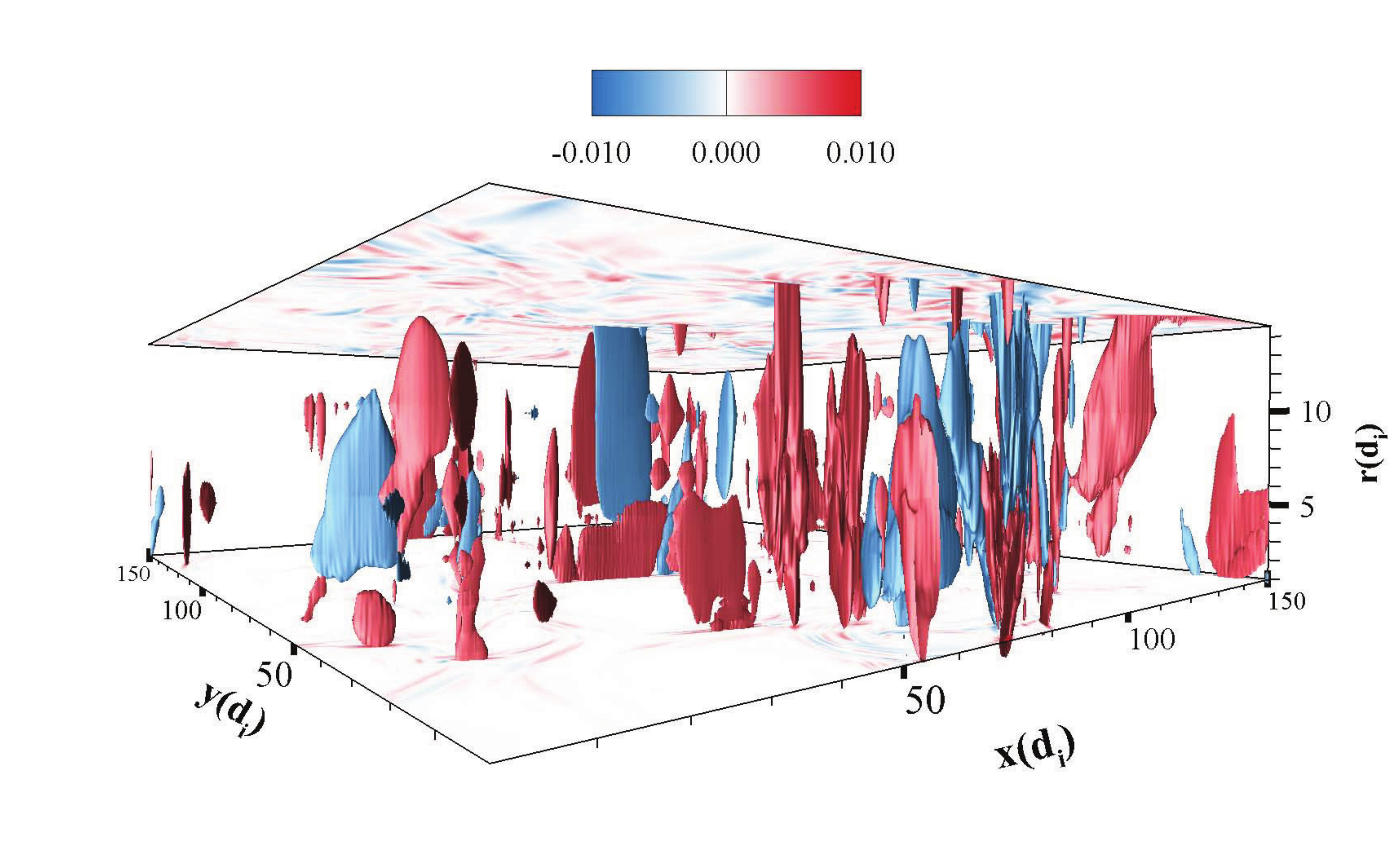}
\caption{Isosurfaces of the LET in a space spanned by the two spatial dimensions $(x,y)$ and
the separation scale $r$.}
\label{Fig.xyz-Yaglom}
\end{figure}

The spatial field of the LET is shown in Fig.~\ref{Fig.xyz-Yaglom}, where we plot isosurfaces of the LET
in a space spanned by the two spatial dimensions $(x,y)$ and
by the separation scale $r$.
The first feature one can see is that the domain is interspersed with positive-negative alternating patches,
and that moreover these patches are such that one would properly call intermittency
since they are localized in real space and broad band in scale.
The LET is signed and its point-wise magnitude could be large but significant cancellations between opposite-signed spots
lead to the global quantity dominated by the forward cascade (positive value).
It is also clear that the structures  are mainly sheet-like -- elongated in the direction of separation length
without tilt, indicating that the location of enhanced energy transfer changes very little as the scale $r$
is varied in the inertial range.

\section{Dominated scales of energy transfer channels}

Elementary manipulations in the full Vlasov-Maxwell system, without reliance on viscous or other closures,
reveal the exchanges between electromagnetic, flow and random kinetic energy \citep{YangEA-PRE-17,YangEA-PoP-17}.
The first three moments of the Vlasov equation, in conjunction with the Maxwell
equations, yield energy equations:
\begin{eqnarray}
\partial_t \mathcal{E}^{f}_\alpha + \nabla \cdot \left( \mathcal{E}^{f}_\alpha \boldsymbol{u}_\alpha + \boldsymbol{P}_\alpha \cdot \boldsymbol{u}_\alpha \right) &=& \left( \boldsymbol{P}_\alpha \cdot \nabla \right) \cdot \boldsymbol{u}_\alpha+ \boldsymbol{j}_\alpha \cdot \boldsymbol{E}, \label{Eq.Ef}\\
\partial_t \mathcal{E}^{th}_\alpha + \nabla \cdot \left( \mathcal{E}^{th}_\alpha \boldsymbol{u}_\alpha + \boldsymbol{h}_\alpha \right) &=& -\left( \boldsymbol{P}_\alpha \cdot \nabla \right) \cdot \boldsymbol{u}_\alpha, \label{Eq.Eth}\\
\partial_t \mathcal{E}^{m} + {\frac{c}{4\pi}} \nabla \cdot \left( \boldsymbol{E} \times \boldsymbol{B} \right) &=& -\boldsymbol{j} \cdot \boldsymbol{E}, \label{Eq.Em}
\end{eqnarray}
where the subscription $\alpha=e, i$ indicates the species, $\mathcal{E}^f_\alpha={\frac{1}{2}}\rho_\alpha \boldsymbol{u}^2_\alpha$ is the fluid flow energy,
$\mathcal{E}^{th}_\alpha={\frac{1}{2}} m_\alpha \int{\left(\boldsymbol{v}-\boldsymbol{u}_\alpha\right)^2 f_\alpha \left(\boldsymbol{x},\boldsymbol{v},t\right) d\boldsymbol{v}}$ is the internal (thermal) energy,
$\mathcal{E}^m={\frac{1}{8\pi}}\left(\boldsymbol{B}^2+\boldsymbol{E}^2\right)$ is the electromagnetic energy,
$\boldsymbol{h}_\alpha$ is the heat flux vector,
$\boldsymbol{j}=\sum_{\alpha} \boldsymbol{j}_\alpha$ is the total electric current density, and
$\boldsymbol{j}_\alpha=n_\alpha q_\alpha \boldsymbol{u}_\alpha$ is the electric current density of species $\alpha$.
This procedure clarifies the roles of several energy transfer channels. For example,
the electromagnetic work, $\boldsymbol{j}\cdot\boldsymbol{E}$, exchanges electromagnetic energy with fluid
flow energy, while the pressure-strain interaction $-\left( \boldsymbol{P}_\alpha \cdot \nabla \right) \cdot \boldsymbol{u}_\alpha$
represents the conversion between fluid flow and internal energy for species $\alpha$.

These energy transfer channels were studied previously in detail \citep{YangEA-PRE-17,YangEA-PoP-17},
with little attempt made to describe their scales of dominance.
Clearly, the plasma turbulence encompasses a vast range of scales and
justification for their dominance at scales, approximate or otherwise,
is crucial. A simple but essential approach to resolve fields
both in space and in scales is the space-filter technique \citep{Germano92},
which, although pervasive in large-eddy simulations, merits more attention
in the plasma turbulence community \citep{YangEA-PoP-17,Camporeale2018coherent}.
The low-pass filtered field of $\boldsymbol{a}(\boldsymbol{x})$,
which only contains information at length scales $>\ell$, is defined as $
\bar{\boldsymbol{a}}_\ell \left(\boldsymbol{x}\right)=\int{d^3 \boldsymbol{r}G_\ell \left(\boldsymbol{r}\right) \boldsymbol{a}\left(\boldsymbol{x}+\boldsymbol{r}\right) }$,
where $\ell$ is the filtering scale,
$G_\ell\left(\boldsymbol{r}\right)=\ell^{-3}G\left(\boldsymbol{r}/\ell\right)$ is a filtering kernel
and $G\left(\boldsymbol{r}\right)$ is a normalized boxcar window function.
To quantify the contribution to the field from different scales,
a scale-band decomposition is introduced as
\begin{equation}
\boldsymbol{a}(\boldsymbol{x})=\sum_n \boldsymbol{a}^{[n]}(\boldsymbol{x}),
\end{equation}
where
\begin{equation}
\boldsymbol{a}^{[n]}(\boldsymbol{x}) = \bar{\boldsymbol{a}}_{\ell_n} \left(\boldsymbol{x}\right)- \bar{\boldsymbol{a}}_{\ell_{n+1}} \left(\boldsymbol{x}\right).
\end{equation}
The band-filtered field $\boldsymbol{a}^{[n]}$ is therefore the fraction of the field $\boldsymbol{a}$ in band $[n]$,
which contains only scales in the band $(\ell_n, \ell_{n+1}]$.
Here these bands are defined with a logarithmic binning $(\gamma^n \ell_0, \gamma^{n+1} \ell_0]$,
where $\gamma>1$ ($\gamma=1.5$ is used in this work) and
$\ell_0$ is taken as the grid spacing of the simulation $\delta x \sim 0.0365 d_i$.
Therefore, the contribution to the pressure-strain interaction and the electromagnetic work
from different scale bands is $-\left( \boldsymbol{P}_\alpha^{[m]} \cdot \nabla \right) \cdot \boldsymbol{u}_\alpha^{[n]}$
and $\boldsymbol{j}^{[m]}\cdot\boldsymbol{E}^{[n]}$, respectively.

\begin{figure}[!htpb]
\centering
\includegraphics[width=0.45\textwidth]{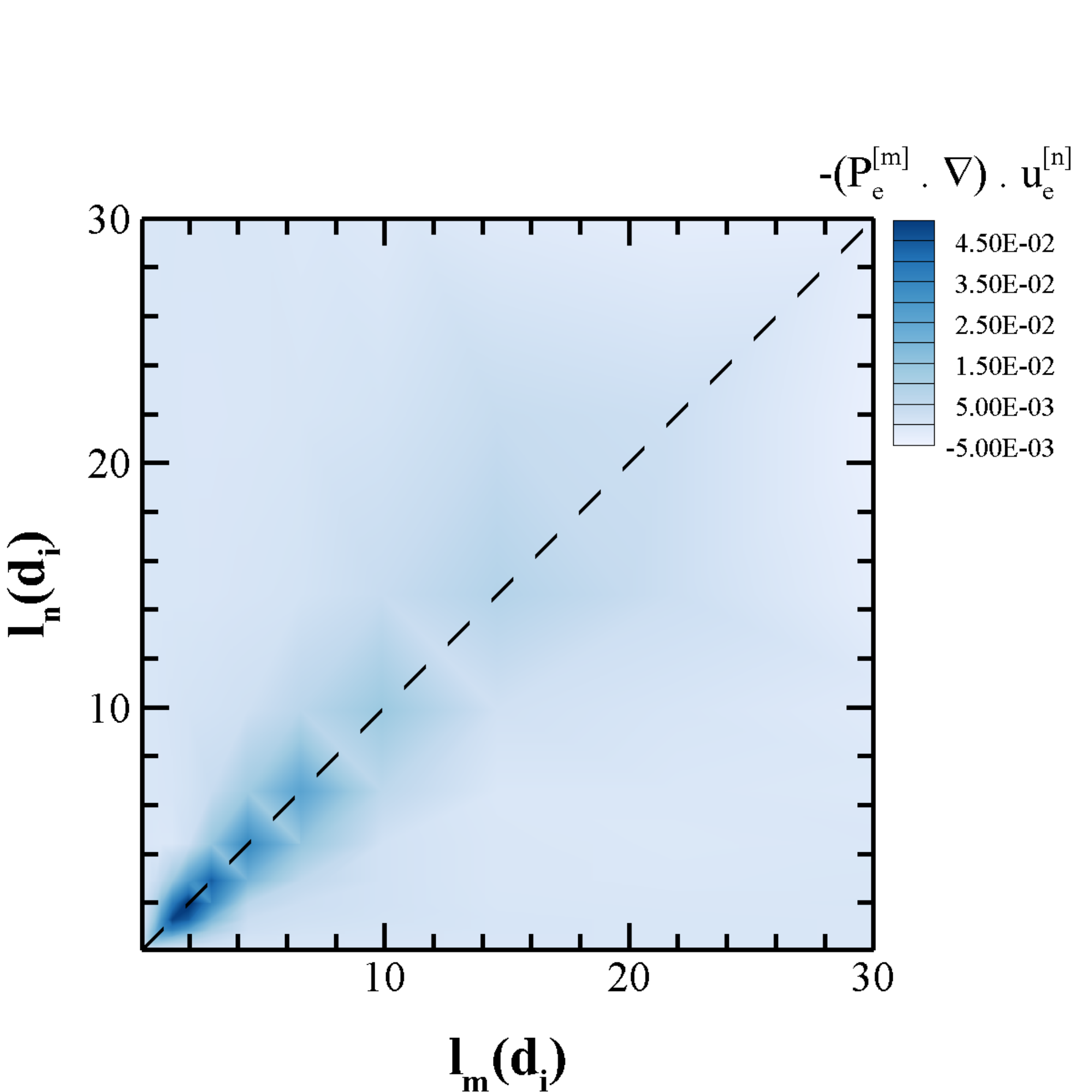}
\includegraphics[width=0.45\textwidth]{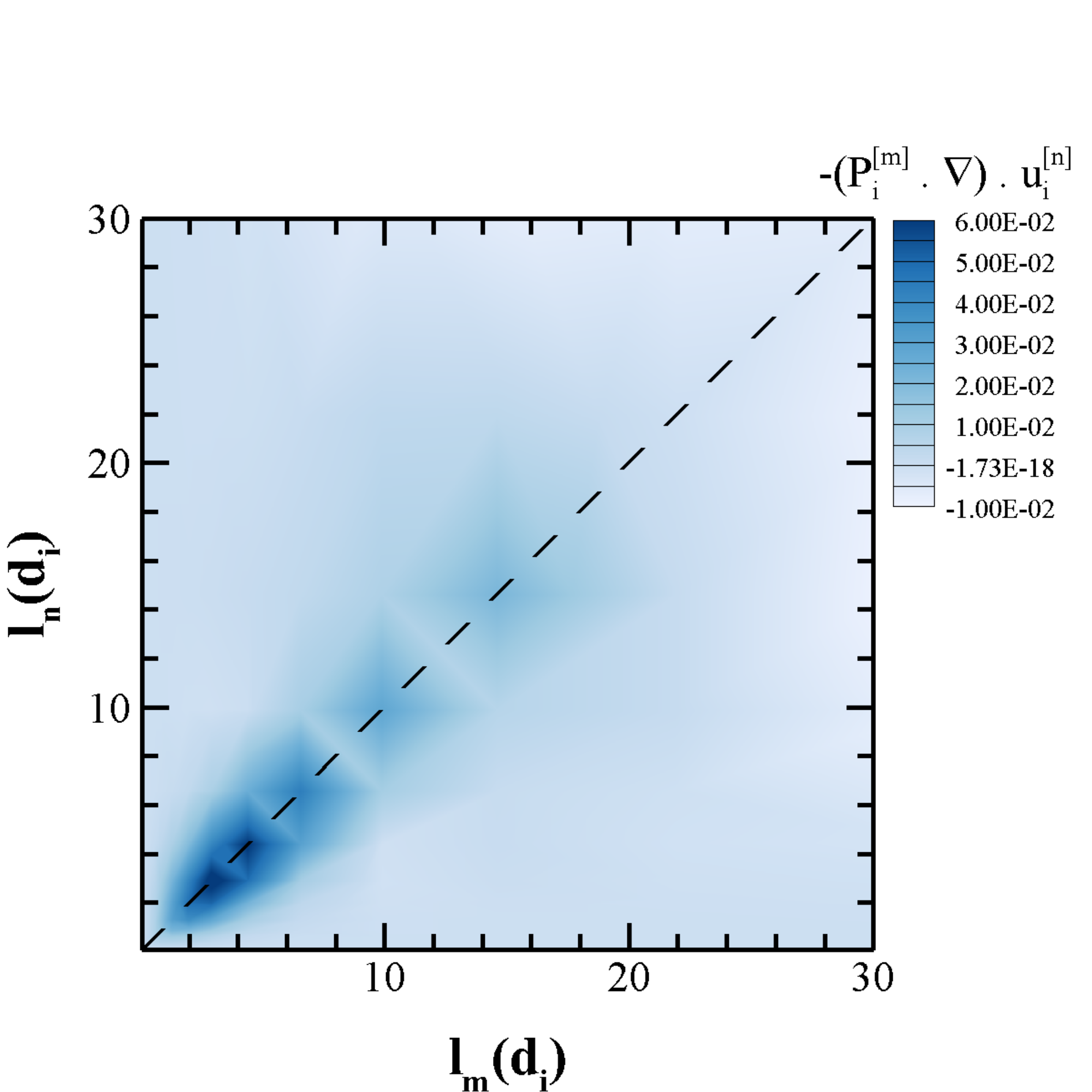}
\caption{Normalized contribution to the pressure-strain interaction from different scale bands,
$\langle -\left( \boldsymbol{P}_\alpha^{[m]} \cdot \nabla \right) \cdot \boldsymbol{u}_\alpha^{[n]} \rangle / \langle -\left( \boldsymbol{P}_\alpha \cdot \nabla \right) \cdot \boldsymbol{u}_\alpha \rangle$,
for both (left) electrons and (right) ions.}
\label{Fig.PS-loc}
\end{figure}

\begin{figure}[!htpb]
\centering
\includegraphics[width=0.45\textwidth]{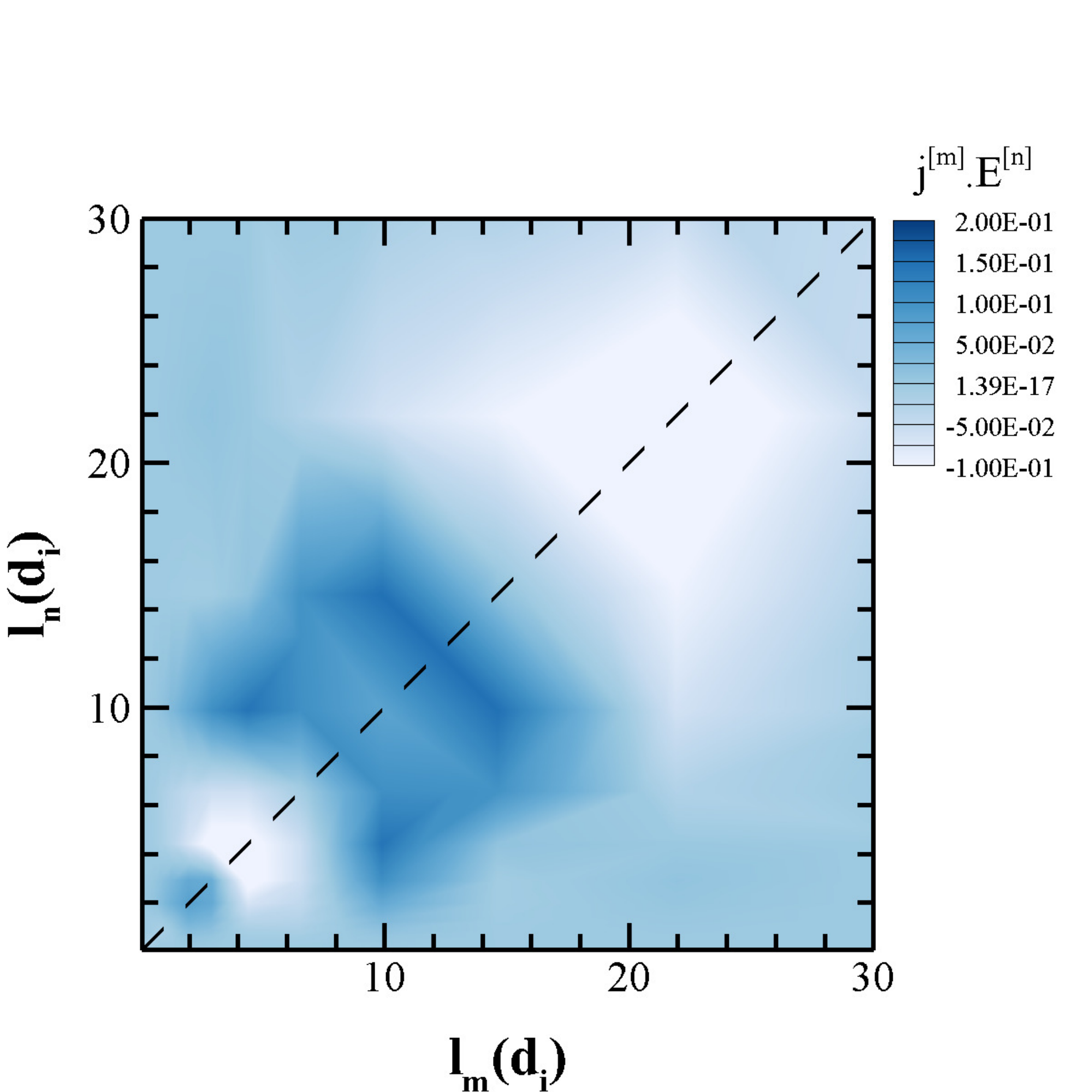}
\caption{Normalized contribution to the electromagnetic work from different scale bands,
$\langle \boldsymbol{j}^{[m]}\cdot\boldsymbol{E}^{[n]} \rangle / \langle \boldsymbol{j}\cdot\boldsymbol{E} \rangle$.}
\label{Fig.EJ-loc}
\end{figure}

Fig.~\ref{Fig.PS-loc} shows the normalized contribution to the pressure-strain interaction
from different scale bands, $\langle -\left( \boldsymbol{P}_\alpha^{[m]} \cdot \nabla \right) \cdot \boldsymbol{u}_\alpha^{[n]} \rangle / \langle -\left( \boldsymbol{P}_\alpha \cdot \nabla \right) \cdot \boldsymbol{u}_\alpha \rangle$, for both electrons and ions. In this paper, the symbol $\langle \cdots \rangle$ denotes a volume average.
Remarkably, the band-filtered pressure-strain interaction for electrons
densely populates along the diagonal as
shown in the plot, as is the case for ions.
They are suggestive of a local interaction, namely, the interaction mainly involves
comparable scales and the contribution from distant bands is negligible.
For the normalized contribution to the electromagnetic work from different scale bands,
$\langle \boldsymbol{j}^{[m]}\cdot\boldsymbol{E}^{[n]} \rangle / \langle \boldsymbol{j}\cdot\boldsymbol{E} \rangle$ in Fig.~\ref{Fig.EJ-loc},
notable departures from the diagonal indicate that a wider range of scales are coupled in the interaction.
It is beyond the scope of this paper to explore in detail the reasons for which
the interactions mainly involve nearby scales, but it may be of some help
to refer to the locality of scale interactions in MHD turbulence
\citep{Verma05,Alexakis05,Teaca11,Aluie10,YangEA-PRE-16}.

Also noteworthy in Fig.~\ref{Fig.PS-loc} is that the most intense (dark blue) contribution
to the pressure-strain interaction is confined to a small region very near the origin, i.e., a few $d_i$.
The contribution to the pressure-strain interaction mostly results from small scales, $<6d_i$ in the present simulation.
Note that the full pressure-strain interaction can be further decomposed as
$-\left(\boldsymbol{P}_{\alpha} \cdot \nabla \right) \cdot \boldsymbol{u}_{\alpha} = -p_{\alpha}\theta_{\alpha}-\Pi_{ij}^{(\alpha)}D_{ij}^{(\alpha)}$,
where $\theta_{\alpha}=\nabla \cdot \boldsymbol{u}_{\alpha}$ is the dilatation,
$D_{ij}^{(\alpha)}=\left(\partial_i u_j^{(\alpha)} + \partial_j u_i^{(\alpha)}\right)/2-\theta_{\alpha} \delta_{ij}/3$
is the traceless strain rate tensor, and
$\Pi_{ij}^{(\alpha)}=P_{ij}^{(\alpha)}-p_{\alpha} \delta_{ij}$ is the deviatoric pressure tensor.
The result shown here poses no contradiction with
the conclusion in \citet{Aluie12,YangEA-PRE-16} that
the pressure-dilatation derives most of its contribution from large scales,
since the pressure-dilatation terms here only account for a small fraction of
$\langle -\left( \boldsymbol{P}_\alpha \cdot \nabla \right) \cdot \boldsymbol{u}_\alpha \rangle$.
For example, the global averages of $-\Pi_{ij}^{(\alpha)}D_{ij}^{(\alpha)}$, $1.05\times 10^{-4}\ {v_{Ar}^3 d_i^{-1}}$ for electrons and
$8.57\times 10^{-5}\ {v_{Ar}^3 d_i^{-1}}$ for ions,
are much greater than those of $-p_{\alpha}\theta_{\alpha}$, $9.03\times 10^{-6}\ {v_{Ar}^3 d_i^{-1}}$ for electrons
and $-8.14\times 10^{-6}\ {v_{Ar}^3 d_i^{-1}}$ for ions.
Therefore, the full pressure-strain interaction behaves quite in analogy with the shear associated part
$-\Pi_{ij}^{(\alpha)}D_{ij}^{(\alpha)}$, which can be cast in viscous dissipation in highly collisional hydrodynamic limit
\citep{Vincenti65,Braginskii65}.
Moving into the realm of strongly compressed plasma, as in the turbulent magnetosheath \citep{Chasapis2018energy},
compressibility effect might make a big difference.

The positive contribution to the electromagnetic work
in Fig.~\ref{Fig.EJ-loc} is concentrated at relatively large scales, $\sim \left[6d_i, 16d_i\right]$
in the present simulation.
While we make no claims of universality of these scale ranges at which
the energy transfer channels dominate,
due to differences in accessible parameters in simulations and in direct observational analysis,
it does qualitatively support the conjecture that
the pressure-strain interaction
mainly operates at small scales,
while the electromagnetic work acts primarily at relatively
large scales.

\section{Spatial correlation between LET and energy transfer channels}

There is accumulating evidence of the association between coherent structures and
plasma dissipation over the last decade.
For example, HVM simulations have shown that
strong distortions of the distribution function occur
near current sheets \citep{Servidio12,Greco12}.
Coherent structures (measured as high ``PVI'' events)
in any of several variables -- density, magnetic field and
velocity, are associated with extremal values of proton
temperature anisotropy \citep{Servidio15}.
One also finds interesting and inter-related roles of vorticity and symmetric strain in
heating \citep{Huba96,DelSarto16,DelSarto2017shear,Franci16,Parashar16}.
Meanwhile, all the energy dissipation proxies discussed above,
the local energy transfer rate (LET),
the electromagnetic work  ${\boldsymbol j} \cdot {\boldsymbol E}$
and the pressure-strain interaction $-\left( \boldsymbol{P} \cdot \nabla \right) \cdot \boldsymbol{u}$,
are systematically concentrated in space, and these concentrations occur within or very near
coherent structures \citep{Sorriso2018local,Sorriso2018statistical,Wan12,Wan16,Parashar16,YangEA-PRE-17,YangEA-PoP-17}.

The connection between coherent structures and energy transfer
represents yet another way
in which coherent structures and intermittency
contribute to plasma dissipation and heating,
further advancing a complementary view that
has been emerging in recent years:
energy cascade leads to several channels of energy conversion, interchange and spatial rearrangement
that collectively leads to production of internal energy.
Given the diversity of energy dissipation proxies that may dominate at different scales,
a significant question is the extent to which they are related
in the overall picture of intermittent heating.

\begin{figure}[!htpb]
\centering
\includegraphics[width=0.22\textwidth]{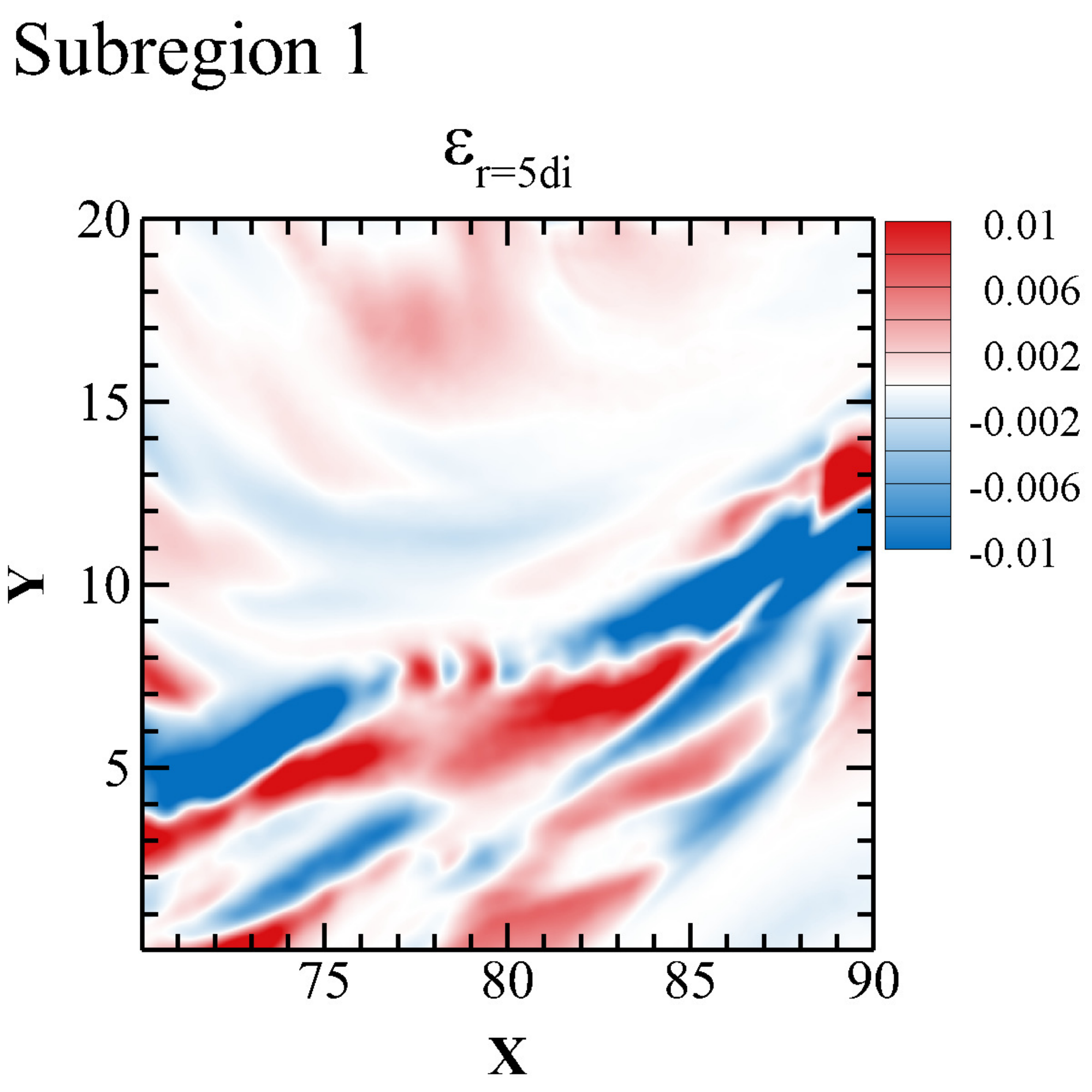}
\includegraphics[width=0.22\textwidth]{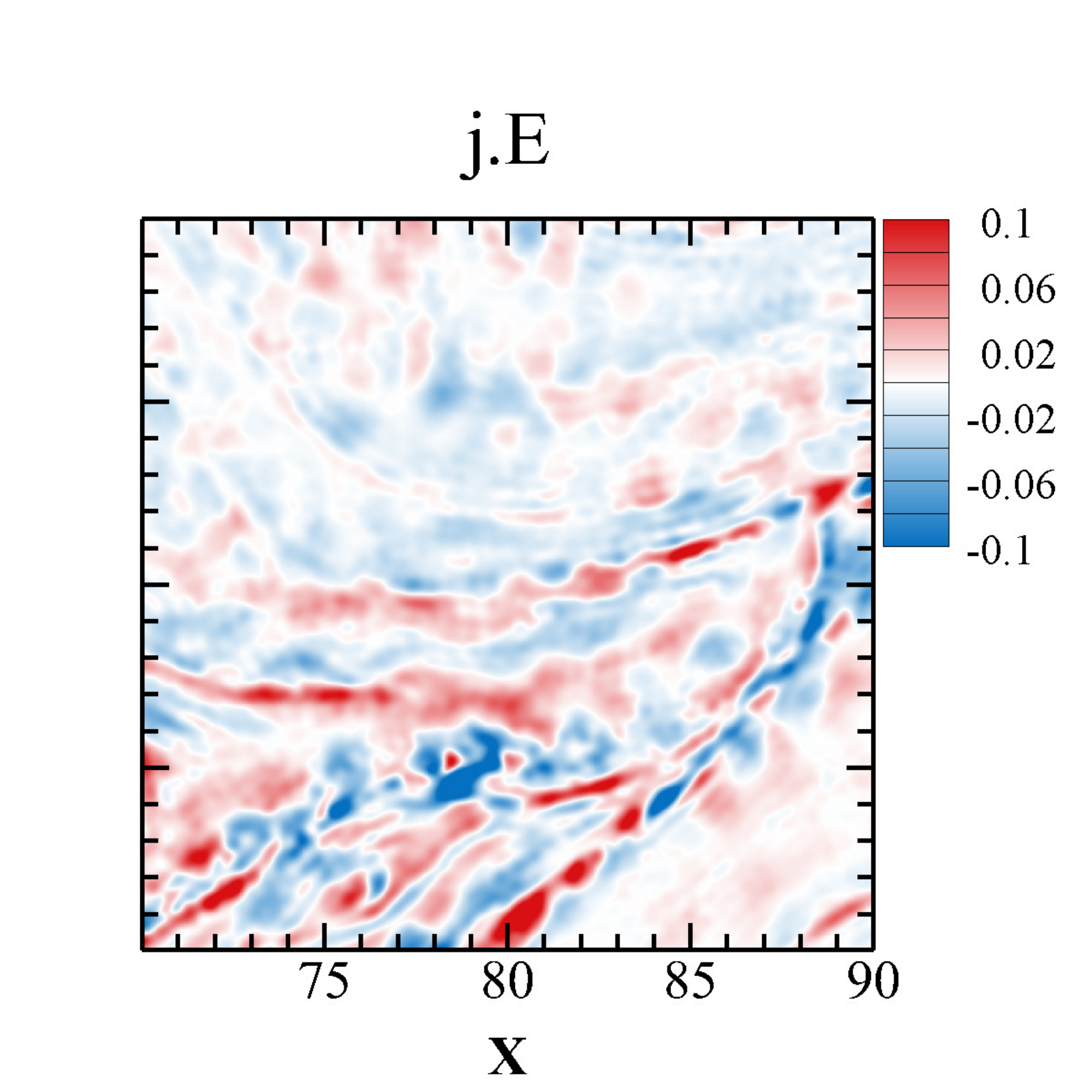}
\includegraphics[width=0.22\textwidth]{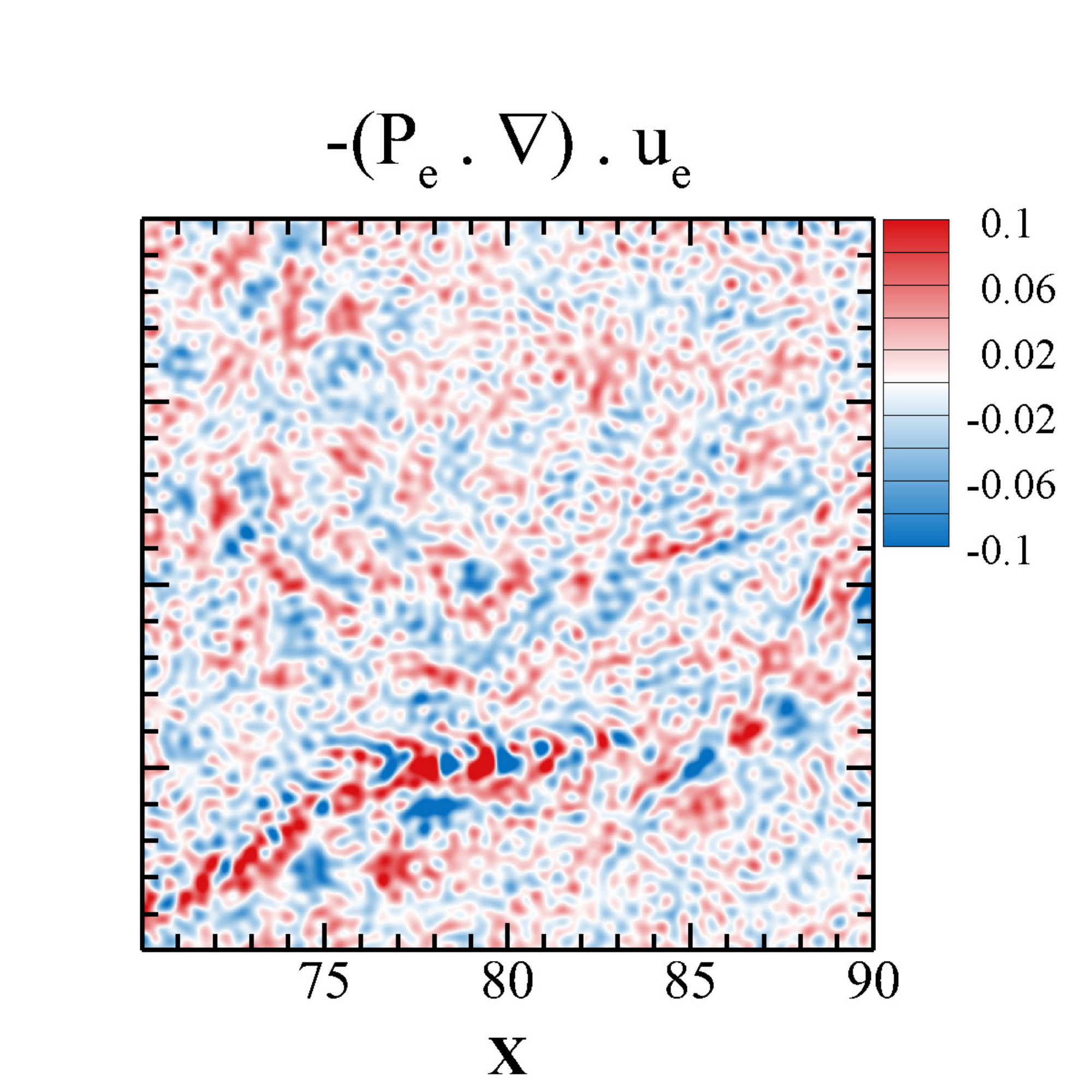}
\includegraphics[width=0.22\textwidth]{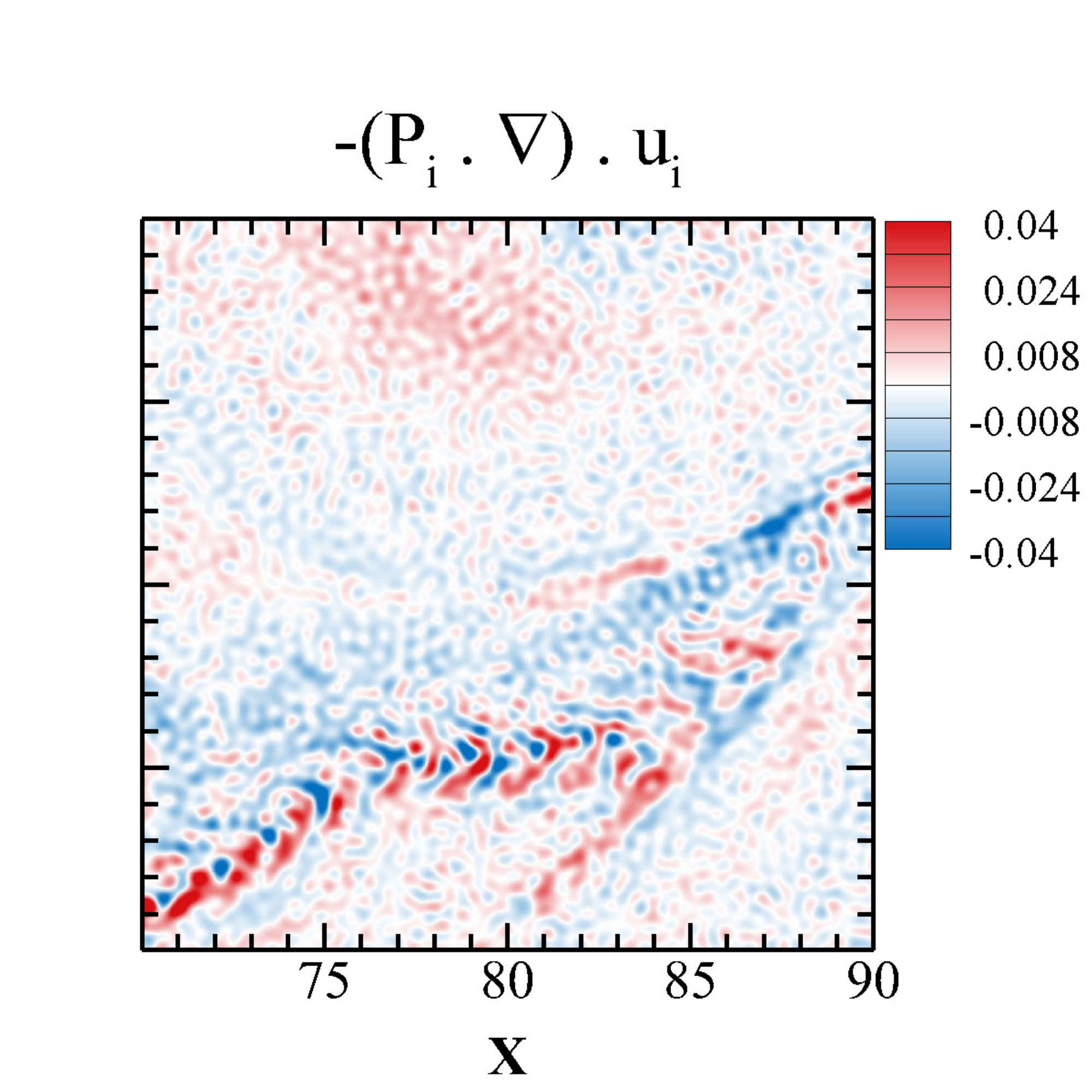}
\includegraphics[width=0.22\textwidth]{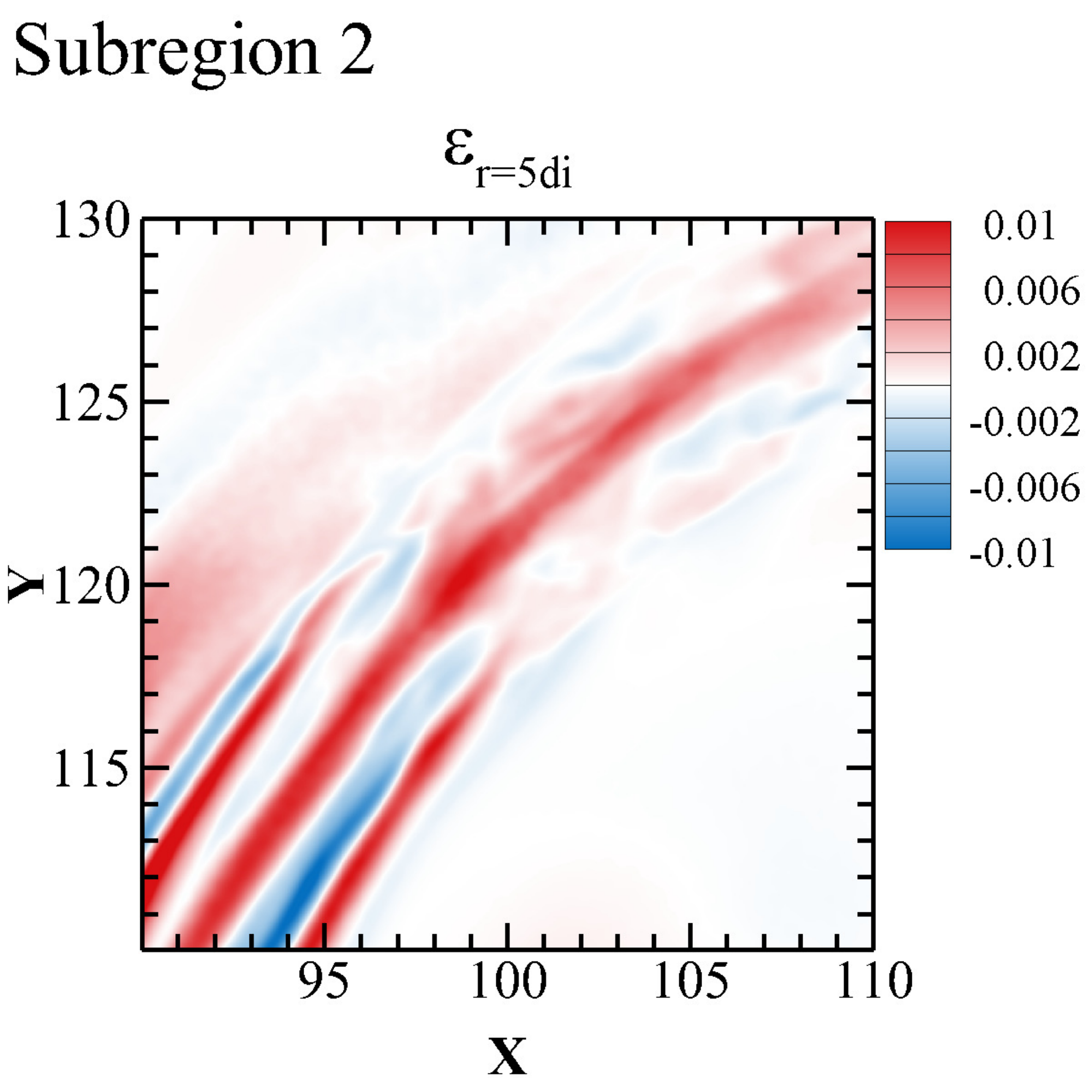}
\includegraphics[width=0.22\textwidth]{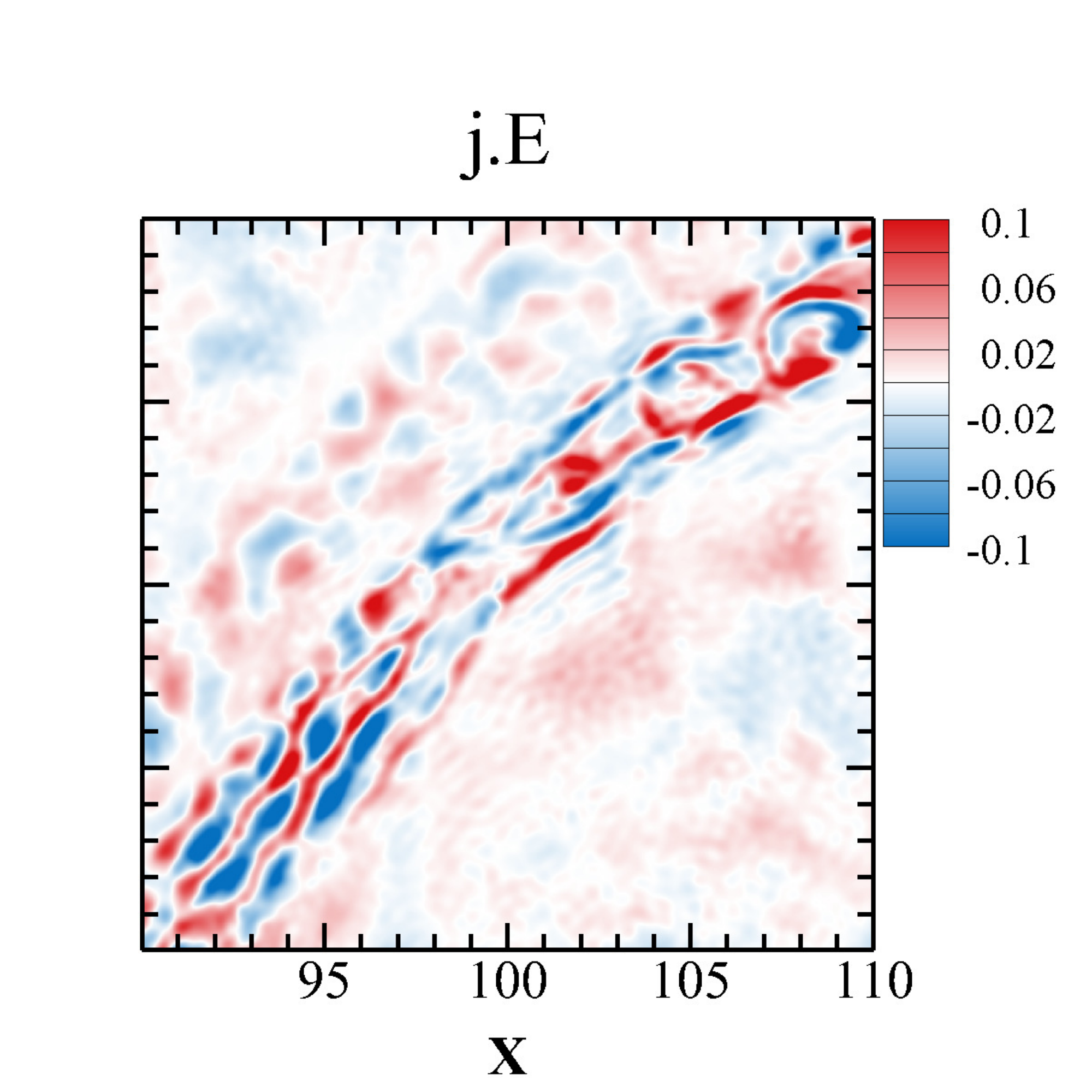}
\includegraphics[width=0.22\textwidth]{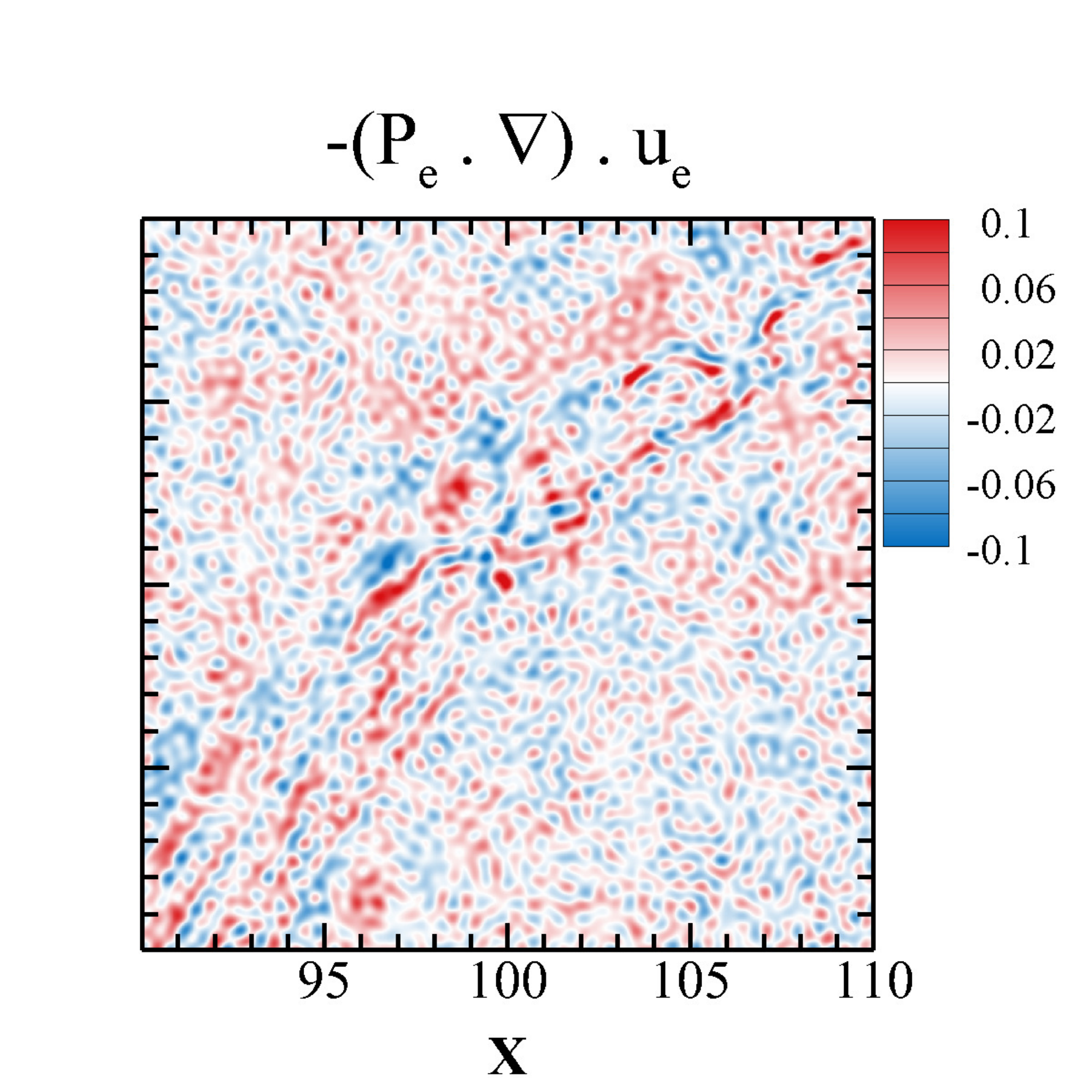}
\includegraphics[width=0.22\textwidth]{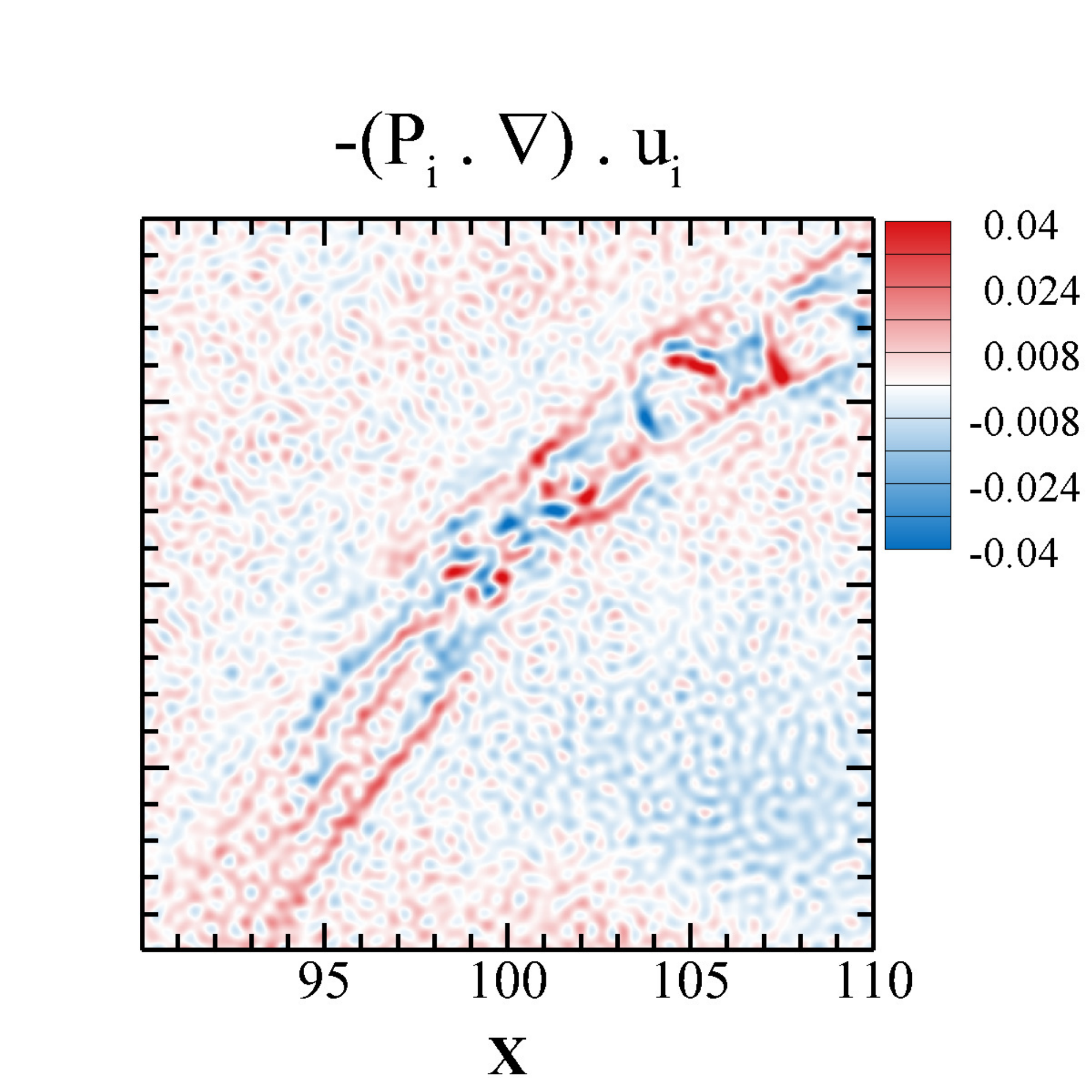}
\includegraphics[width=0.22\textwidth]{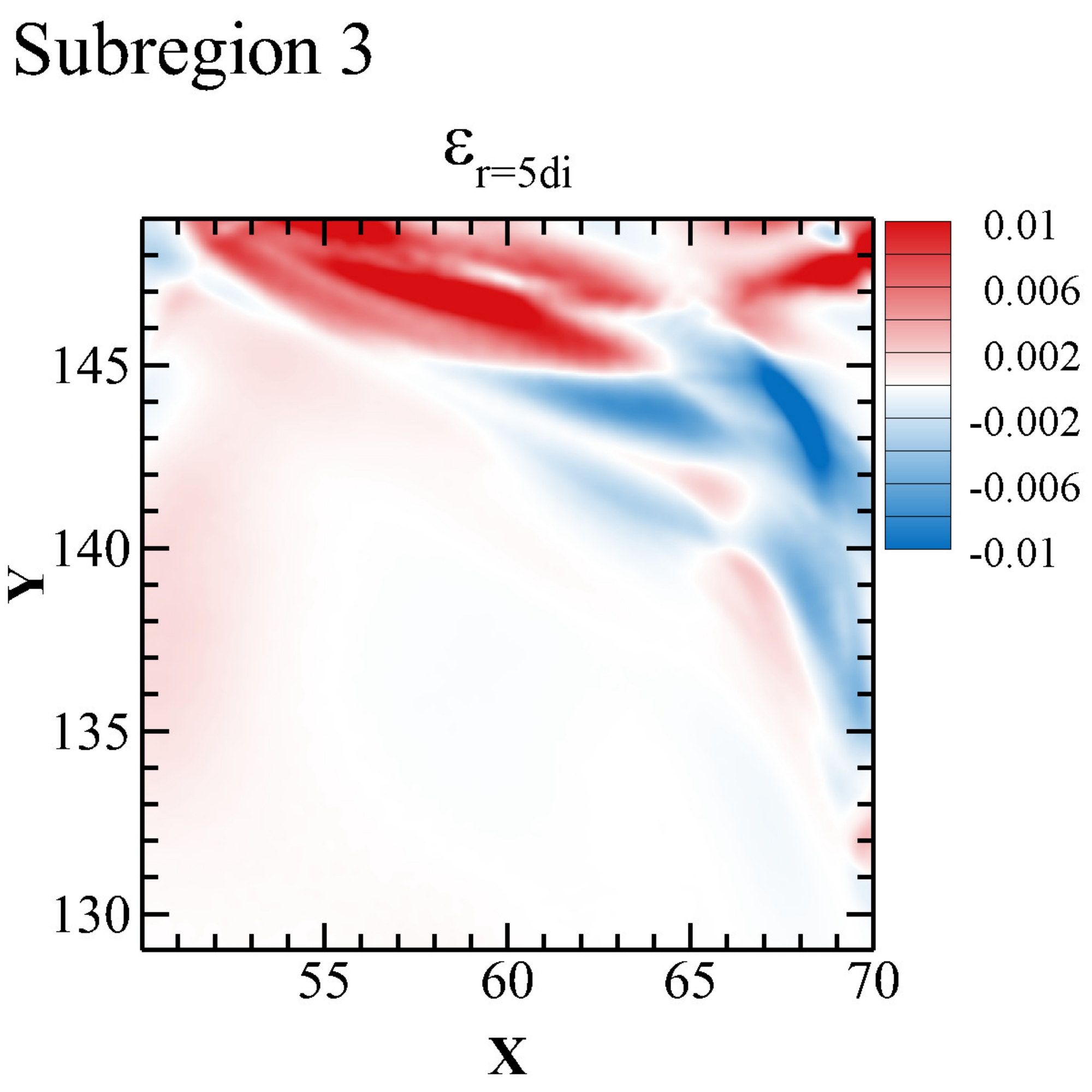}
\includegraphics[width=0.22\textwidth]{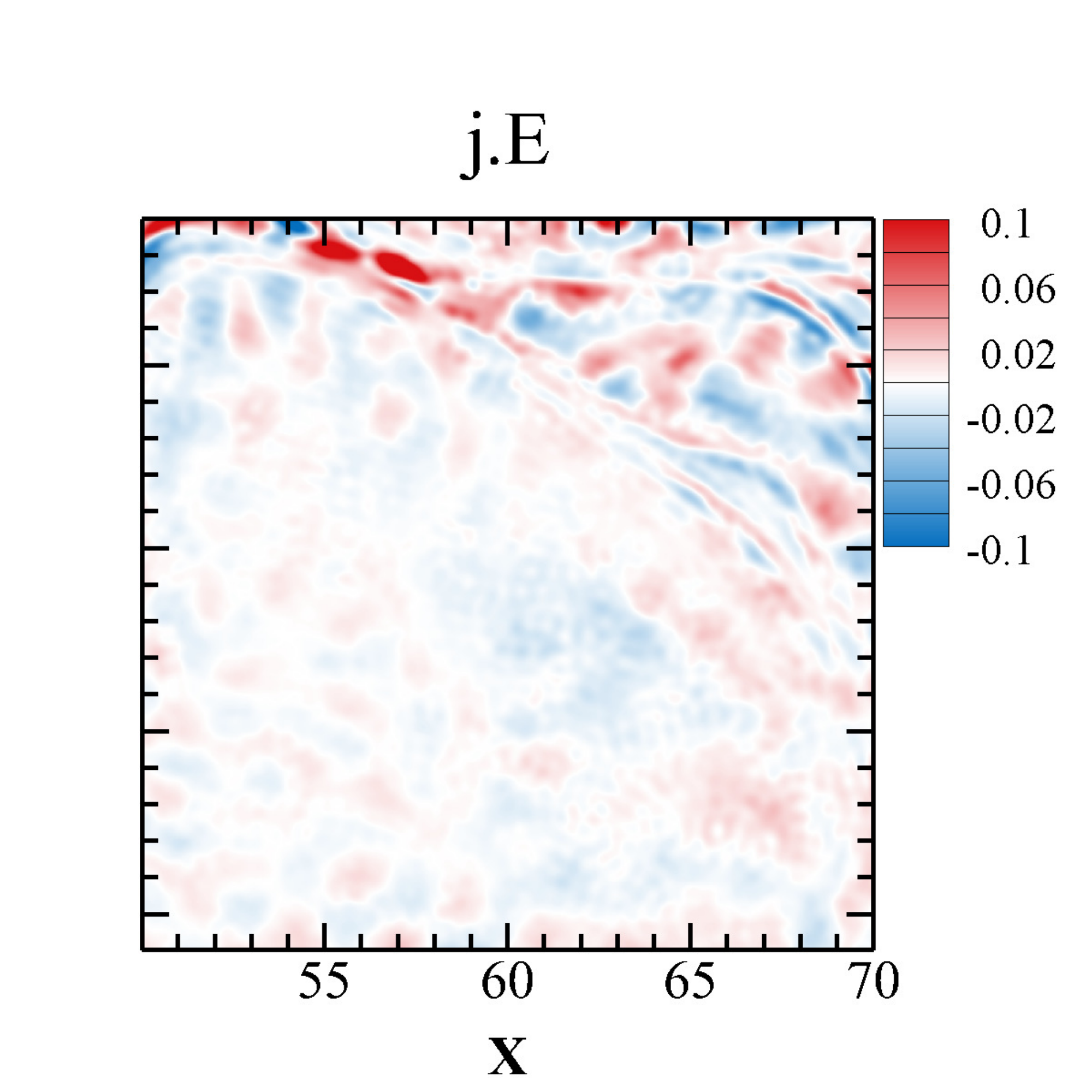}
\includegraphics[width=0.22\textwidth]{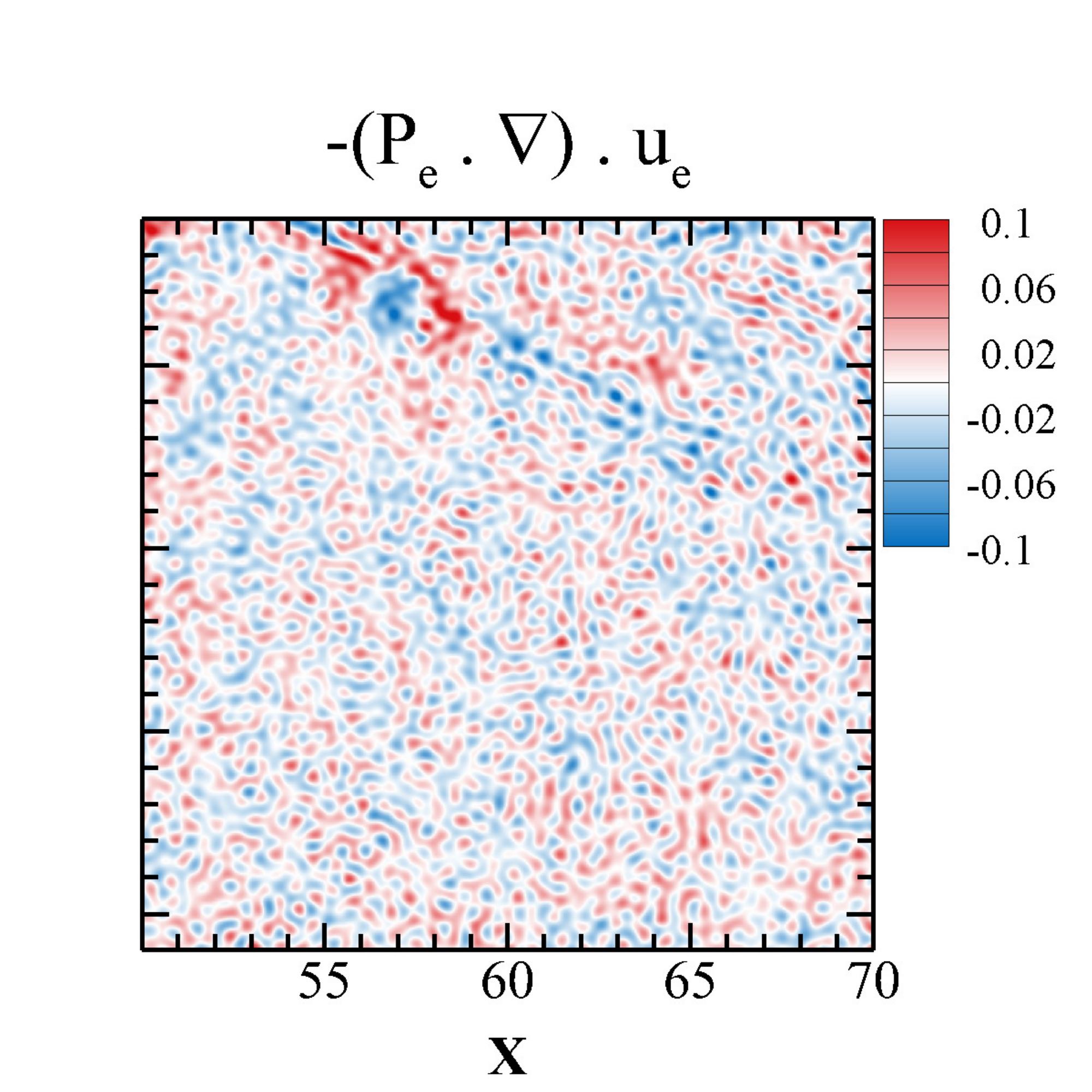}
\includegraphics[width=0.22\textwidth]{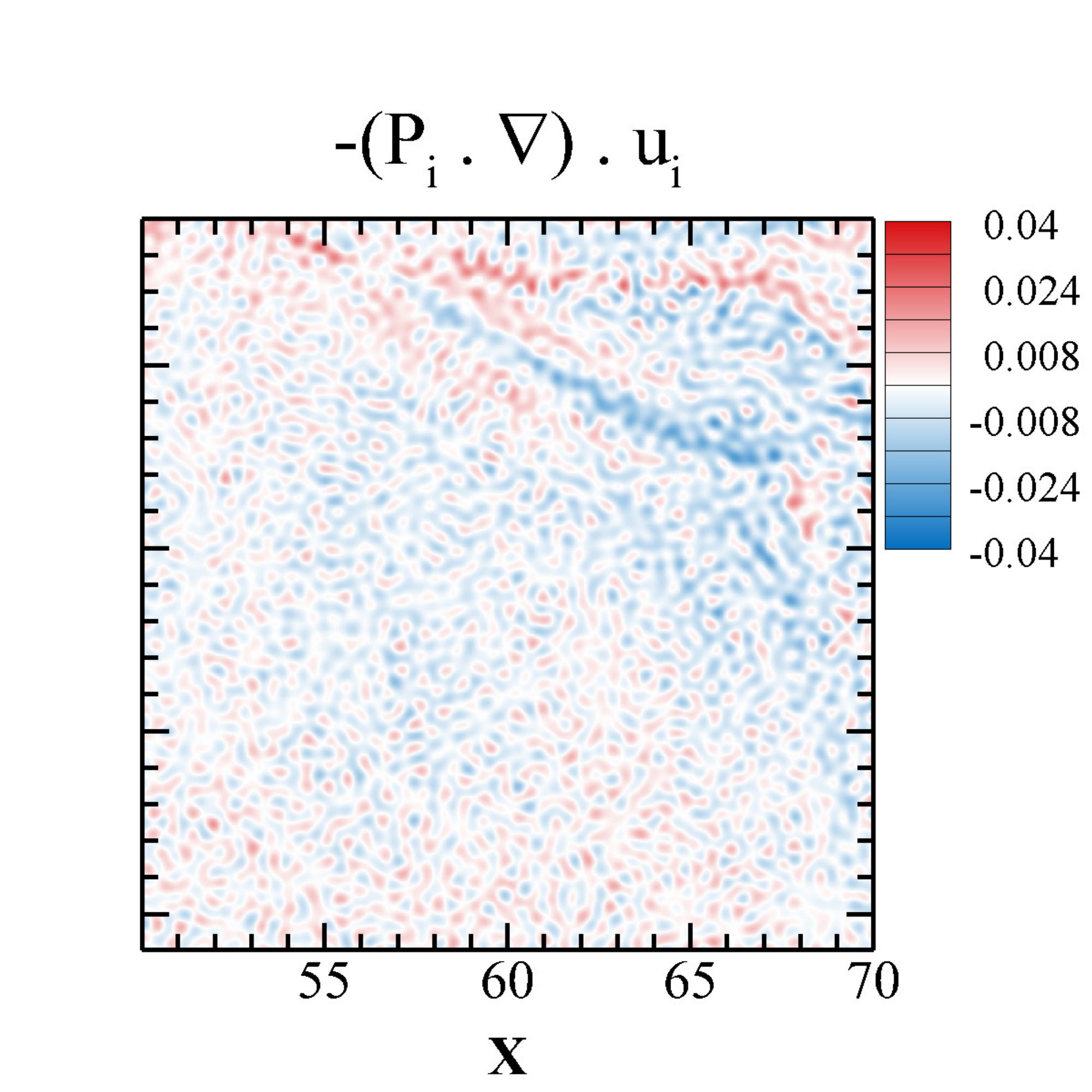}
\caption{Contours for $\epsilon_{r=5d_i}$,
$\boldsymbol{j} \cdot \boldsymbol{E}$,
$-\left( \boldsymbol{P}_e \cdot \nabla \right) \cdot \boldsymbol{u}_e$ and
$-\left( \boldsymbol{P}_i \cdot \nabla \right) \cdot \boldsymbol{u}_i$
at small subregions.}
\label{Fig.disp-EJ-PD}
\end{figure}

For a first diagnostic to address these questions,
Fig.~\ref{Fig.disp-EJ-PD} shows spatial contour maps of the LET at $5d_i$ (i.e., $\epsilon_{r=5d_i}$),
$\boldsymbol{j}\cdot\boldsymbol{E}$ and
$-\left( \boldsymbol{P} \cdot \nabla \right) \cdot \boldsymbol{u}$ separately for protons and electrons
in several subregions.
The first thing to notice is that these energy dissipation proxies are highly localized with
intense values concentrated at small regions.
Also seen immediately is the greatly similar pattern of their spatial distributions,
though point-wise magnitudes and signs might be different.
The striking similarity between the spatial patches of the different proxies
suggests that coherent structures, energy transfer and
dissipation are all correlated to a certain extent.

We compute the global volume averages of energy dissipation proxies, as shown in Table~\ref{Table.avr}.
These entries in the Table, wherein $\langle -\left( \boldsymbol{P}_e \cdot \nabla \right) \cdot \boldsymbol{u}_e \rangle + \langle -\left( \boldsymbol{P}_i \cdot \nabla \right) \cdot \boldsymbol{u}_i \rangle \sim 1.92\times 10^{-4}\ ({v_{Ar}^3 d_i^{-1}})$,
are meaningfully compared with the mean energy dissipation rate estimated by the third-order law,
$\epsilon \sim 1.87 \times 10^{-4}\ ({v_{Ar}^3 d_i^{-1}})$.
It is interesting to note that $\langle\boldsymbol{j} \cdot \boldsymbol{E}\rangle$ is negative, suggesting that kinetic energy of particles
is converted into electromagnetic energy at this moment.
We found that the global average of the electromagnetic work
oscillates significantly over time at high frequencies (comparable to $\omega_{pe}$).
This is likely an artefact of artificial value of $\omega_{pe}/\omega_{ce}$ in our simulation,
and could be remedied by time averaging the results over a plasma oscillation period \citep{HaggertyEA17}.
As computed by \citet{Wan15,Wan16}, the energy conversion rate in the frame moving with electrons 
$\langle D_e \rangle =\langle \boldsymbol{j} \cdot \left(\boldsymbol{E}+\boldsymbol{u}_e \times \boldsymbol{B}\right) - \rho_c \left(\boldsymbol{u}_e \cdot \boldsymbol{E}\right) \rangle$ is 
somewhat lower than $\epsilon$ and 
$\langle -\left( \boldsymbol{P}_e \cdot \nabla \right) \cdot \boldsymbol{u}_e \rangle + \langle -\left( \boldsymbol{P}_i \cdot \nabla \right) \cdot \boldsymbol{u}_i \rangle$.
Thus this measure may only account for a part of total dissipation.

\begin{table}[!htpb]
\centering
\begin{threeparttable}
\caption{Volume averages of the LET, the electromagnetic work, the pressure-strain interaction
and the electron-frame dissipation measure $D_e$ \citep{Zenitani2011new}.
Quantities listed are in the code units ${v_{Ar}^3 d_i^{-1}}$.}
\label{Table.avr}
\begin{tabular}{lcccccc}
\hline
 $\epsilon$ \tnote{a} & $\langle\boldsymbol{j} \cdot \boldsymbol{E}\rangle$ & $\langle \boldsymbol{j}_e \cdot \boldsymbol{E} \rangle$ \tnote{b} & $\langle\boldsymbol{j}_i \cdot \boldsymbol{E}\rangle$ \tnote{c} & $\langle -\left( \boldsymbol{P}_e \cdot \nabla \right) \cdot \boldsymbol{u}_e \rangle$  & $\langle -\left( \boldsymbol{P}_i \cdot \nabla \right) \cdot \boldsymbol{u}_i \rangle$ & $\langle D_e \rangle$ \tnote{d} \\
\hline

 $1.87\times 10^{-4}$ & $-5.1 \times 10^{-5}$ & $1.24 \times 10^{-4}$ & $-1.75\times 10^{-4}$ & $1.14\times 10^{-4}$ & $7.8\times 10^{-5}$ & $1.42\times 10^{-4}$ \\
 \hline
\end{tabular}
\begin{tablenotes}
 \item[a] Volume average of the LET $\epsilon_r$ evaluated within $\left[2d_i, 10d_i\right]$ range.
 \item[b,c] Separate contributions of electrons and protons to the electromagnetic work.
 \item[d] The work done by electromagnetic fields on particles, evaluated in the frame of 
 electron bulk motion.
\end{tablenotes}
\end{threeparttable}
\end{table}

More diagnostics, such as scatter plots of any two proxies (not shown here) and
the corresponding Spearman correlation coefficients $\rho_s$,
can be used to clarify the possible correlation. However, their Spearman correlation coefficients
are rather small, e.g., $\rho_s(\left|\epsilon_{r=5d_i}\right|, \left|\boldsymbol{j} \cdot \boldsymbol{E}\right|)=0.25$,
$\rho_s(\left|\epsilon_{r=5d_i}\right|, \left|-\left( \boldsymbol{P}_e \cdot \nabla \right) \cdot \boldsymbol{u}_e\right|)=0.042$ and
$\rho_s(\left|\boldsymbol{j} \cdot \boldsymbol{E}\right|, \left|-\left( \boldsymbol{P}_e \cdot \nabla \right) \cdot \boldsymbol{u}_e\right|)=0.057$.
The conclusion drawn in this way may seem at first to be in conflict with the finding of Fig.~\ref{Fig.disp-EJ-PD}.
But it is maybe not so surprising that there
is not a strong point-wise correlation amongst
the LET, the pressure-strain interaction and the electromagnetic work.
This recalls the lower correlation between proton heating and
current relative to vorticity, and related findings \citep{Servidio15, DelSarto16, Franci16, Parashar16}.
According to Eqs.~\ref{Eq.Ef},\ref{Eq.Eth} and \ref{Eq.Em}, the transport terms on the left-hand side
could be locally enormous, thus spoiling co-location of the proxies.

\begin{figure}[!htpb]
\centering
\includegraphics[width=0.6\textwidth]{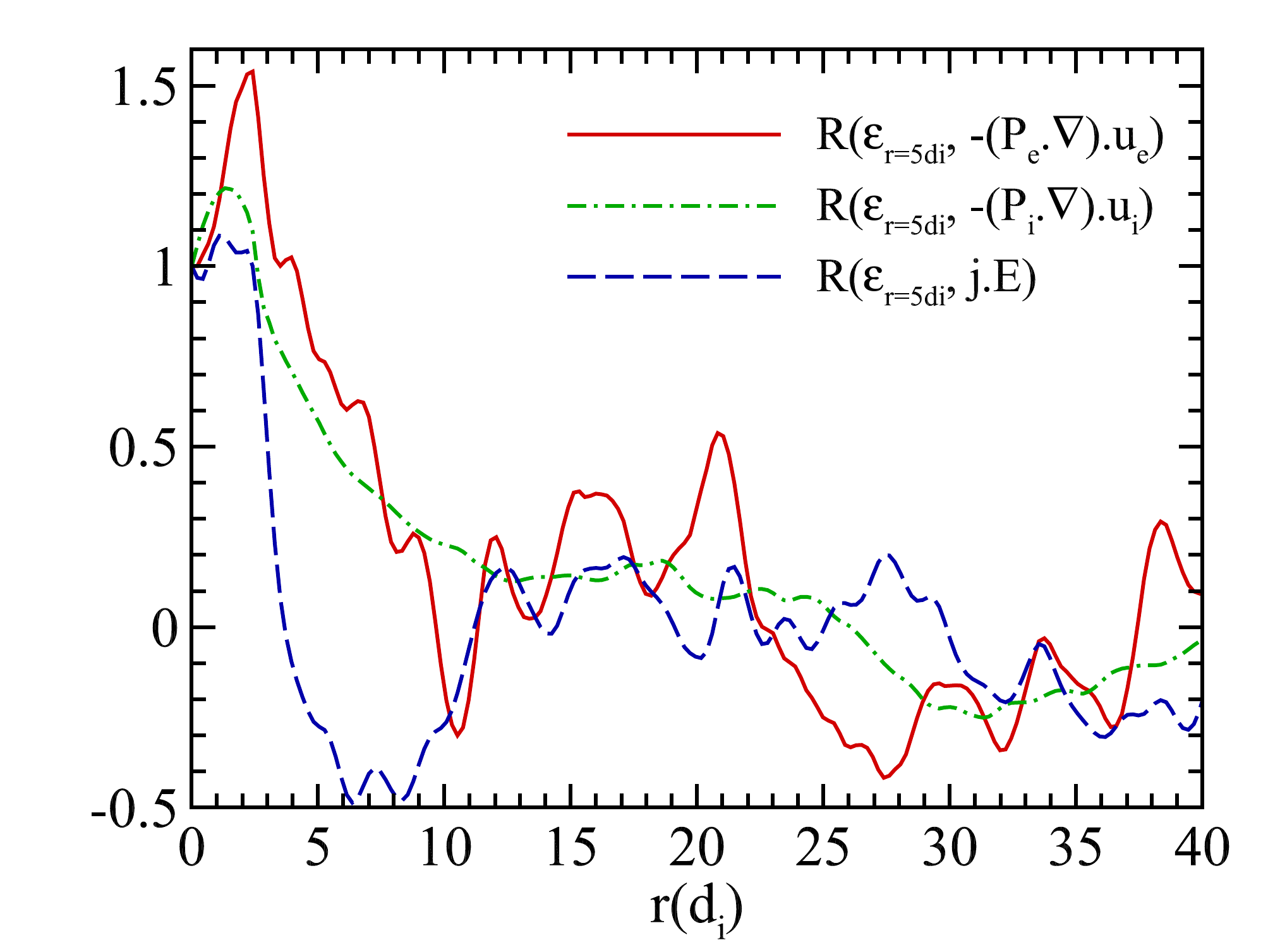}
\caption{Two-point cross correlation functions of $\epsilon_{r=5d_i}$,
$\boldsymbol{j} \cdot \boldsymbol{E}$,
$-\left( \boldsymbol{P}_e \cdot \nabla \right) \cdot \boldsymbol{u}_e$ and
$-\left( \boldsymbol{P}_i \cdot \nabla \right) \cdot \boldsymbol{u}_i$.}
\label{Fig.CrossCor}
\end{figure}

We anticipate then that the energy dissipation proxies in the subsequent energy transfer
are juxtaposed to, but not exactly co-located with one another.
This complex spatial arrangement can best be illustrated by
a scale-dependent cross-correlation function,
\begin{equation}
R(f,g,r)=\frac{\langle \left(f(\boldsymbol{x}+\boldsymbol{r})-\langle f\rangle\right) \left(g(\boldsymbol{x})-\langle g\rangle\right) \rangle}{\langle \left(f(\boldsymbol{x})-\langle f\rangle\right) \left(g(\boldsymbol{x})-\langle g\rangle\right) \rangle},
\end{equation}
where $f$ and $g$ are the fields and
the direction of displacement $\boldsymbol{r}$ is arbitrary for isotropic turbulence in the plane.
Such correlations have been used to show strong correlations between hotter plasma and vorticity
as compared to the correlations between hotter plasma and current \citep{Parashar16},
although the correlations were not normalized in the same fashion there.
Seen in Fig.~\ref{Fig.CrossCor} is that the correlation curves peak near $2d_i$.
It is natural that the statistical connection between the proxies will
become infinitely attenuated as the points become infinitely far apart in space
(This follows from the familiar clustering property of turbulent fluctuations).
It is then possible to calculate a correlation length, $\lambda_c(f,g)=\int_0^{\infty} R(f,g,r) dr$,
a convenient measure of the spatial extent over which the fields are appreciably correlated.
The results are shown in Table~\ref{Table.CorLength}.
There is clear delocalization between the proxies, so that
$-\left( \boldsymbol{P}_\alpha \cdot \nabla \right) \cdot \boldsymbol{u}_\alpha$
and $\boldsymbol{j} \cdot \boldsymbol{E}$ are larger near, not necessarily at,
locations of large $\epsilon_{r=5d_i}$.
Note that the short correlation length associated with $\boldsymbol{j} \cdot \boldsymbol{E}$
is in part due to its reversal of sign around spatial separation $r=4d_i$.

\begin{table}[!htpb]
\centering
\caption{Correlation lengths estimated from the data in Fig.~\ref{Fig.CrossCor}.}
\label{Table.CorLength}
\begin{tabular}{lc}
\hline
  Proxies  & $\lambda_c$ ($d_i$)   \\
\hline
$\left(\epsilon_{r=5d_i}, -\left( \boldsymbol{P}_e \cdot \nabla \right) \cdot \boldsymbol{u}_e \right)$  & 7.85 \\
$\left(\epsilon_{r=5d_i}, -\left( \boldsymbol{P}_i  \cdot \nabla \right) \cdot \boldsymbol{u}_i  \right)$  & 4.72 \\
$\left(\epsilon_{r=5d_i}, \boldsymbol{j} \cdot \boldsymbol{E} \right)$   & 1.77 \\
 \hline
\end{tabular}
\end{table}

\section{Conclusions}

The dissipative mechanism in weakly collisional plasma is a topic that pervades decades of studies without a
consensus solution.
One popular approach to explain dissipation is to resort to wave particle interactions,
with instabilities regulating the dynamics of extreme distortions of the distribution function.
A complementary view is that several channels of energy conversion emerge in a turbulence cascade process.
In this paper, we study energy dissipation proxies based on the cascade, i.e.,
the local energy transfer rate (LET), the electromagnetic work $\boldsymbol{j} \cdot \boldsymbol{E}$ and
the pressure-strain interaction $-\left( \boldsymbol{P}_\alpha \cdot \nabla \right) \cdot \boldsymbol{u}_\alpha$.

We find that although these proxies are displaced in space,
enhanced electromagnetic work and pressure-strain interaction are concentrated
in the proximity of regions with intense LET.
Their connection is a somewhat atypical property in that
it is not readily found using point-wise correlation,
but rather one must appeal to statistics, such as
two-point cross correlation functions,
to understand it.
The basis for their association but also delocalization in space lies in
a recognition of the key steps of energy transfer:
conservative rearrangement of energy in space due to transport terms;
conservative rearrangement of energy in scales due to energy cascade;
electromagnetic work on particles that drives flows;
and pressure-strain interactions that produces internal energy.

The association between the LET and energy transfer channels in this paper,
in conjunction with the results in Hall MHD \citep{Camporeale2018coherent}
and compressible MHD \citep{YangEA-PRE-16,YangEA-POF-17},
should be an adequate starting point for further investigating,
for example, how do the characteristics of energy transfer vary going from MHD to kinetic scales.
We show here that
there is a decade of range, $\sim \left[2d_i, 10d_i\right]$, over which the third-order law is valid.
It is also found that
contributions to $\boldsymbol{j} \cdot \boldsymbol{E}$ and
$-\left( \boldsymbol{P}_\alpha \cdot \nabla \right) \cdot \boldsymbol{u}_\alpha$ are mainly from
large ($\sim \left[6d_i, 16d_i\right]$ in this simulation) and small ($<6d_i$ in this simulation)
scales, respectively. Therefore, these proxies are dominated at different scales.

Taken together, their connections and differences further support this intuitive picture \citep{YangEA-PRE-17}:
the cascade drives scale-to-scale energy transfer, with a net transfer of energy to small scales, and
leads to intermittent distributions of several channels of energy conversion
that in turn provide the dominate dissipation mechanism. Electromagnetic energy is converted into flows by
electromagnetic work, while pressure-strain interaction converts energy from flows into internal energy.
Note that none of the three dissipation proxies we examined are sign-definite, and to therefore
some type of averaging is necessary to interpret any of them as a net rate of conversion or transfer
in the complex pathways to dissipation and heating.
It is worth emphasizing that our 2.5D PIC simulation is not intended to reproduce
any particular solar wind feature or data interval.
While this model is a powerful tool, it also fails to properly account for important real effects,
such as solar wind expansion, three dimensionality,
ion-to-electron mass ratio, and so on.
Consequently, a report such as the present one necessarily
leads to an incomplete description of energy dissipation.
We anticipate our results to be extended to more sophisticated models in future works.

\begin{acknowledgments}
This work has been supported by NSFC Grant
Nos. 91752201, 11672123, and 91752000; the Thousand
Talents Plan for Young Professionals; the Shenzhen Science
and Technology Innovation Committee (Grant No. JCYJ20170412151759222).
This research was partially supported by the University of Delaware.
\end{acknowledgments}



%

\end{document}